\documentclass[pdflatex,sn-chicago]{sn-jnl}%

\usepackage{graphicx}%
\usepackage{multirow}%
\usepackage{amsmath,amssymb,amsfonts}%
\usepackage{amsthm}%
\usepackage{mathrsfs}%
\usepackage{stmaryrd}%
\SetSymbolFont{stmry}{bold}{U}{stmry}{m}{n} 
\usepackage{bm}
\usepackage[title]{appendix}%
\usepackage{xcolor}%
\usepackage{tcolorbox} 
\usepackage{textcomp}%
\usepackage{manyfoot}%
\usepackage{booktabs}%
\usepackage{algorithm}%
\usepackage{algorithmicx}%
\usepackage{algpseudocode}%
\usepackage{listings}%
\usepackage{caption}
\usepackage{subcaption}
\usepackage{anyfontsize}
\usepackage{hyperref}

\begin{document}

\title[Article Title]{Many Retrocausal Worlds: A Foundation for Quantum Probability}

\author{Michael Ridley}
\email{mikeridleyphysics@gmail.com}
\affil{Faculty of Engineering and Institute of Nanotechnology and Advanced Materials,
Bar Ilan University, Ramat Gan 52900, Israel}
\affil{School of Physics and Astronomy, Tel Aviv University, Tel Aviv, 69978, Israel}


\abstract{Recent accounts of probability in the many worlds interpretation of quantum mechanics are vulnerable due to their dependence on probability theory per se. For this reason, the many worlds interpretation continues to suffer from the incoherence and quantitative problems. After discussing various theories of probability, I discuss the incoherence problem and argue that self-locating probabilities centered in time-extended worlds can solve it. I then discuss and refute various solutions to the quantitative problem. I argue that the only tenable way to ground these self-locating probabilities is to identify the mathematical form of the Born rule as a generic pattern in a time-extended wavefunction, and to distribute degrees of belief over the region of wavefunction occupied by this pattern. I then outline a time-symmetric version of quantum mechanics - the Fixed Point Formulation - which, interpreted within a time-symmetric Everettian framework, can provide the foundation for a theory of quantum probability.}


\maketitle

\tableofcontents
\section{Introduction and Overview}
\subsection{Posing the problem}
The measurement problem of textbook nonrelativistic quantum mechanics refers to one of several tensions between the four basic postulates of this theory:

\begin{itemize}
    \item[] \textbf{P1} The ontological postulate - physical systems are represented at each moment of time by a state vector in Hilbert space, $\left|\Psi\left(t\right)\right\rangle$
    
    \item[] \textbf{P2} The dynamical postulate - the state vector of a system described by Hamiltonian operator $\hat{H}$ evolves unitarily in accordance with the time-dependent Schrödinger equation between measurements
    \begin{align}\label{eq:TDSE}
    \centering
    i\hbar\frac{\partial\left|\Psi\left(t\right)\right\rangle }{\partial t}  =  \hat{H}\left|\Psi\left(t\right)\right\rangle 
    \end{align}
    \item[] \textbf{P3} The composition postulate - a state composed of distinct parts A and B corresponding to distinct subsystems is a tensor product state in the composite Hilbert space $\mathcal{H}_{A+B}=\mathcal{H}_{A}\otimes\mathcal{H}_{B}$
    
    \item[] \textbf{P4} The statistical postulate - when a measurement takes place at time $t$, the state vector `collapses' into one of several possible states $\left\{ \left|\phi_{i}\right\rangle \right\}$  with a probability given by the Born rule (\cite{born1926quantenmechanik}): 
    \begin{align}\label{eq:Born}
    \centering
        Pr\left(i,t\right)=\left|\left\langle \phi_{i}\right.\left|\Psi\left(t\right)\right\rangle \right|^{2}
    \end{align}
    
\end{itemize}

The source of conceptual difficulty is usually located in \textbf{P4}. It is easy to see why - the character of the statistical postulate is quite unlike those of \textbf{P1-P3}, which describe a physical system evolving in accordance with a deterministic, linear and reversible dynamical rule. The ontological and dynamical postulates \textbf{P1}, \textbf{P2} exist in logical harmony with each other and play analogous roles to those of particle coordinates and laws of motion, respectively, in Newtonian mechanics. The composition postulate \textbf{P3} accounts for perhaps the most peculiar aspect of quantum theory - entanglement (\cite{schrodinger1935discussion}). These postulates make no reference whatsoever to the intervention of an `observer'. They certainly do not describe a nomologically privileged observer whose experience can only be explained with a novel law of evolution and a metaphysical commitment to chance. So it appears that \textbf{P4} contains both physical and metaphysical elements which are not present in \textbf{P1-P3}.

We may follow \cite{maudlin_three_1995} and categorize versions of the measurement problem by their emphasis on the different clauses in \textbf{P4}:

Firstly, there is the \textbf{problem of unique outcomes} - the unobserved wavefunction evolves unitarily into a superposition of physically distinct states, but only one of these is actually observed during measurement. 

Secondly, there is the \textbf{problem of random collapse} - the mechanism by which a distinct state is realized during observation is apparently stochastic, contrary to the deterministic evolution in time of the unobserved state. This metaphysical inconsistency becomes a logical one if the measurement time is not fixed: in regions of time where a measurement may have occurred or not, there are two contradictory laws describing the system evolution (\cite{bell_against_1990,rovelli_incerto_1998}). Moreover, there is no direct experimental evidence whatsoever for collapse in nature (\cite{vaidman_schizophrenic_1998,vaidman_many-worlds_2021,vaidman_probability_2012}). 

Thirdly, there is the \textbf{quantitative problem} - can the mathematical form of Eq. \eqref{eq:Born} be derived from postulates \textbf{P1-P3}? If not, what additional assumptions can be used to derive it? What kind of assumption could serve as sufficient grounds for choosing one derivation over another? One may also ask the question outright: \emph{why} derive the Born rule? What can be gleaned from a derivation? 

\cite{hartle2021we} has made the case that one should value derivations of the Born measure to the extent that they tell us about \emph{connections} in the theory. They should tell us what are the main ingredients in our fundamental representation of nature, such that, in an explication of quantum probability, something is \emph{learned} which enhances the \textbf{generalizability} of the theory. This is done when strong mathematical assumptions are replaced with weaker assumptions, reduced in number or defined in such a way that they explicate the meaning of the main ingredients of the theory so they find a comfortable place in our general conception of reality.

\subsection{Realism}

Any theory which attempts to reconcile the tensions between \textbf{P1-P4} must make explicit basic assumptions about the source of the observed randomness in physics, and must explain the observed relative frequencies in terms of these assumptions. It is moreover important to keep in mind that the theoretical framework of quantum mechanics arose from the need to interpret the statistical data from experiments on the interaction of matter and light at atomic lengthscales (\cite{shankar2012principles}). Historically, it was the success of this theory in predicting the relative frequencies of outcomes in those type of experiments which caused it to replace classical mechanics as a calculational tool. Such considerations, in my view, mean that the quantum theory is obviously realist, in the sense that there is an external world which it represents in some non-trivial sense via the ontological postulate \textbf{P1} (\cite{albert2013wave}). The reason for this particular choice of representation has always been its astonishing empirical success.

There is, moreover, a fundamental conceptual distinction between the ontology of a physical system and the epistemic situation of an observer who may or may not be interacting with that system. Metaphysics deals with generic structural aspects of the world, for instance whether or not the laws of nature are deterministic. It provides a picture of reality that grounds the truth of propositions that can be made about it, in the following sense:

\emph{Ontology explains experience - our experiences are instantiations of physical properties on ontological structure.}

Since ontology is a representation of reality which explains our experience, and experience is the source of our knowledge, ontology \emph{explains} our knowledge. This is connected to \textbf{generalizability} of our explanatory framework - an ontology which predicts individual experiences is a more logically parsimonious theory than one which lists them. As \cite{berenstain_privileged-perspective_2020} has argued, the underlying representation itself takes a `bird's-eye' view of nature, acknowledging the fact that there are many observers and describing them equivalently. I characterize the latter position as follows:

\textbf{Neutral realism} - quantum mechanics describes reality in a perspective-neutral way.

There are many competing realist interpretations which replace the basic picture of a single vector in Hilbert space evolving unitarily. 

Some of them modify \textbf{P1} so that the ontology of the theory contains fundamental entities besides the wavefunction. This is the route taken by the mechanics of deBroglie-Bohm (dBB) \cite{bohm_suggested_1952}, which treats the wavefunction as a field on high-dimensional space `guiding' the emergent three-dimensional structure of everyday experience \cite{albert_elementary_1996}. 

Other realist approaches modify the dynamical postulate \textbf{P2} to explicitly include a stochastic dynamical element. This is the strategy of physical collapse theories such as the Ghirardi-Rimini-Weber (GRW) spontaneous collapse theory (\cite{ghirardi_unified_1986}), the continuous spontaneous localisation theory of \cite{pearle_combining_1989} and the gravity-induced wavefunction collapse models of \cite{diosi1987universal,penrose1998quantum}. Such theories involve an explicit metaphysical commitment to indeterminism, as the unitary dynamics of standard quantum theory is replaced by nonlinear dynamics and the consequent breakdown of quantum superpositions (\cite{bassi_models_2013}). However, both the dBB pilot wave theory and physical collapse theories run into significant, perhaps insurmountable complications when combined with special relativity (\cite{durr_can_2014,tumulka_relativistic_2006}). 

\subsection{Probability and its observers}

In an early discussion of probability in quantum mechanics, \cite{feynman_concept_1951} repeatedly emphasized that it was the new calculus of probability which lay at the heart of quantum physics:

\begin{quote}
\emph{“The new theory asserts that there are experiments for which the exact outcome is fundamentally unpredictable, and that in these cases one has to be satisfied with computing probabilities of various outcomes. But far more fundamental was the discovery that in nature the laws of combining probabilities were not those of the classical probability theory of Laplace.”}
\end{quote}

Reading this, one could be forgiven for viewing the problem of probability in quantum mechanics as a purely mathematical endeavour, a logic relating “measurements” in abstracto. This is the route taken by the quantum logic program (\cite{pitowsky_quantum_2006}). However there is a huge explanatory gap in theories which use this program to claim resolution of the measurement problem. Firstly, it suffers from an absence of truthmakers - external physical structures which \emph{explain} experience (\cite{huber2023defending}).  

Secondly, measurements are physical processes, describing sequences of events which involve transfer of charge, energy and matter between spacetime regions. It is this material arena that we wish to describe and which ultimately grounds the quantum theory. \cite{feynman_concept_1951} goes on to make precisely this point:

\begin{quote}
\emph{“We and our measuring instruments are part of Nature and so are in principle described by an amplitude function satisfying a deterministic equation. Why can I only predict the probability that a given experiment will lead to a definite result? From whence does the uncertainty arise?”}
\end{quote}

Clearly, the Copenhagen orthodoxy was not acceptable to Feynman, who seems here to require from physics something beyond mere rules of prediction - he is indicating the possibility of a unified framework that explained the experience of randomness in an experiment on the basis of a deeper deterministic theory. 

It is also clear that, for Feynman, the system-observer separation was not consistent with a basic realist stance on the content of the theory. As he then goes on to say (\cite{feynman_concept_1951}):

\begin{quote}
    \emph{"The usual separation of observer and observed which is now needed in analyzing measurements in quantum mechanics should not really be necessary, or at least should be even more thoroughly analyzed."}
\end{quote}

The simultaneous treatment of observer and observed is required to formulate quantum cosmology - a bird's-eye representation of the universe which can be generalized to include the outcomes of experiments conducted at all times and spatial locations (\cite{berenstain_privileged-perspective_2020}). Attaining this representation is the proper object of unification in theoretical physics. Indeed, a bird's-eye point of view seems to me a necessary prerequisite for any interpretation of quantum mechanics which hopes to survive unification with general relativity. If there is an indexical element to certain observations depending on local aspects of a measurement, then this must also be included in the theory, as well as the rules for translating between different localized points of view. The broader framework includes them all, and explains them in terms of a plausible interpretation of the laws. In the context of quantum mechanics, \cite{saunders_decoherence_1993} and \cite{wallace_worlds_2002} contend that this should look like a theory which moves from the foundational postulates \textbf{P1-P3} (which are concerned with the bird's-eye view) and predict the local experience of observers from them, much like one can use Minkowski spacetime structure and the Lorentz transformations to predict observations made in a single foliation of spacetime. There is moreover a logical virtue in theoretical parsimony - the less independent mathematical statements required to reconcile different theories, the less the number of propositions a unified theory has to be consistent with. 

\subsection{The MWI}\label{MWI}

Such considerations lead \cite{everett_relative_1957} to take the view that any observer $O$ who is physically connected to the system $S$ being measured obeys the same physical laws as $S$ and therefore $O$ does not possess a nomologically privileged perspective on reality. $O$ is a lump of physical material interacting with the physical matter in $S$, and many different $O$'s may be having similar experiences. The combined state of system plus observer is represented by a vector in Hilbert space in accordance with \textbf{P1} and \textbf{P3}, and evolved unitarily in accordance with \textbf{P2}, where the observer-system interaction is included in the Hamiltonian of Eq. \eqref{eq:TDSE}. The combined state of all subsystems in the universe is called the universal wavefunction $\left|\Psi_{U}\right\rangle$  (\cite{everett_relative_1957,everett_iii_theory_1973}) - it is the only object needed to model reality in the theory. Therefore the logical parsimony of Everett's formulation is accompanied by a certain parsimony of ontology, which I henceforth refer to as \textbf{wavefunction monism} (\cite{vaidman2016all,calosi_quantum_2018}). 

Without \textbf{P4}, the quantum theory has the same meta-logical structure of classical mechanics: there is one postulate stating what there is, and another stating how it changes. It changes unitarily, and there is no non-linear collapse process in the theory. Thus, \cite{vaidman_quantum_2014} argues, determinism is restored to physics. However, the presence of an interaction term in the Hamiltonian during measurement implies that the combined $S+O$ state branches into an entangled superposition of wavevectors, each of which describes a distinct physical situation for both the observer and the system they observe. The interpretative step taken by Everett at this point was to simply accept what the mathematics is telling us at face value: there are many versions of the observer which exist on parallel branches of reality (\cite{everett_relative_1957}). These branches are sometimes called `worlds' - despite disagreement in the literature about what exactly constitutes a world (\cite{wilson_everettian_2012,wilson_objective_2013,cunningham_branches_2014,marchildon_multiplicity_2015}), this formal branching into a superposition of distinct macroscopic states provides the basic logic underlying the many worlds interpretation (MWI) of quantum mechanics (\cite{vaidman_many-worlds_2021}).

We can cash out \textbf{wavefunction monism} in two directions:

\textbf{Complete neutral realism} - every pattern in nature can be mapped to a structure in the wavefunction in a perspective-neutral way.

The assumption of \textbf{complete neutral realism} is nothing less than the assumption that quantum mechanics is a general framework for the description of nature. It is impossible, for instance to even begin to start doing quantum cosmology and constructing theories of quantum gravity without this assumption.  

The acceptance of the reality of parallel branches in the wavefunction is not strictly speaking a consequence of \textbf{complete neutral realism} - one could allow that all entities in nature are representable in the wavefunction without allowing all structures in the wavefunction to be representative of reality. The reverse implication must therefore be specified: 

\textbf{Extensive neutral realism} - every structure in the wavefunction can be mapped to a pattern in nature in a perspective-neutral way.

The type of realism employed in the MWI is the union of \textbf{complete neutral realism} with \textbf{extensive neutral realism}:

\textbf{Strong neutral realism} - there is a one-to-one mapping between structures in the wavefunction and patterns in nature which preserves perspective neutrality.

That is to say, discussion of reality or of the wavefunction refers fundamentally to the same thing. Observers are just parts of reality, and are therefore just parts of the wavefunction. All moments of time are represented in the wavefunction equivalently, and so therefore are all experiments. In the context of the MWI, \textbf{wavefunction monism} and \textbf{strong neutral realism} are identical metaphysical theses. 

The MWI contains two parts - the core structure of the theory, and a picture of that structure which makes sense of physical experience (\cite{vaidman2022many}). After having identified the observer with the observer state, a split in the observer state implies a split of the observer. This branching process solves the \textbf{problem of random collapse} - a direct interpretation of unitary evolution without wavepacket reduction implies a sudden division of the observer into many different `centers of subjectivity' on different branches of the wavefunction distinguished by the different measurement outcomes (\cite{albert_probability_2010,wilson_objective_2013}). It also solves the \textbf{problem of unique outcomes}, effectively by denying that there are unique outcomes - everything which can occur does occur, but physically distinct occurrences correspond to distinct branches of the wavefunction. This removes the problem of action at a distance from the theory - distinct regions of the wavefunction are nonseparable, belonging to a single underlying substance - the wavefunction - which grounds the phenomena of our everyday experience (\cite{miller_quantum_2016,ismael_quantum_2020}).

As \cite{saunders2021everett} has pointed out, the MWI must come with an account of what probabilities are, since it excludes non-unitary evolution from the outset. This means it must both acknowledge and draw from the centuries-old debate over the nature of chance. It must also explicate quantum probability in a way that is fit for the quantum purpose. Two major difficulties emerge for the Everettian account of probability (\cite{vaidman2022many}):

\textbf{The incoherence problem} 

In the MWI, it seems that all outcomes of an experiment occur with probability 1. There does not seem to be a meaning for uncertainty about the future within this deterministic framework. However, there is the clear appearance of objective chance at work in physical systems. 

\textbf{The quantitative problem}

This is, as stated above, the problem of justifying or deriving the mathematical form of the Born probability measure from the structural postulates \textbf{P1-P3}. 

There have been many attempts to do this - for a full review of approaches within different interpretations, see \cite{vaidman_derivations_2020}. In the context of the MWI, the dominant argument strategy of recent times stems from the work of Deutsch and Wallace, who ask us to assign probabilities based on the decisions made by a rational agent in a quantum universe (\cite{deutsch_quantum_1999,wallace_emergent_2012}). Derivations of the Born measure based on the exploitation of physical symmetries in quantum systems can be found in \cite{vaidman_probability_2012,mcqueen_defence_2019,short2023probability}, and an argument from a different kind of epistemic symmetry is used for the same purpose in \cite{carroll_many_2014,sebens_self-locating_2018}. More generally, \cite{barrett_typical_2017} makes the case that in \emph{any} version of the MWI, one has a "special obligation" to make explicit the auxiliary assumptions regarding probabilities, precisely because of the lack of any probabilistic element within the pure unitary mechanics of wavefunctions.

The MWI is an interpretation which follows a \emph{monistic} representation of reality by the wavefunction. The structure of the universal wavefunction is assumed to be in one-to-one correspondence with the structure of the universe, a bird's-eye map of reality which can be brought to bear on the local experience of an observer on a particular wavefunction branch, explaining their experience. In his emphasis on cosmological support for wavefunction realism, Everett was followed by \cite{dewitt_quantum_1970} and \cite{susskind_copenhagen_2016}, amongst others. However, there are several different versions of wavefunction realism within the MWI, depending on which essential aspects of reality are included in the specification of the theory (\cite{chen_realism_2019}). These include formulations which treat $\left|\Psi_{U}\right\rangle$  as an aggregation of decoherent time-extended histories (\cite{saunders_many_2010,saunders2021everett}) and the spacetime state realism of \cite{wallace_quantum_2010} which constructs $\left|\Psi_{U}\right\rangle$  from connected regions of spacetime. It also includes the novel reductionist program of \cite{carroll_mad-dog_2019} which attempts the derivation of spacetime structure from abstract Hilbert space structure alone. Each of these versions of the MWI includes a version of the postulates \textbf{P1-P3}. However they all retain the core characteristics of those postulates and the Everettian picture of nature - realism about a single highly-structured vector in Hilbert space evolving unitarily. 

Because of its logical simplicity, \cite{blackshaw2023probability} argues that a direct realist take on the core quantum theory is generalizable to quantum theories that go beyond nonrelativistic quantum mechanics - the basic structure of unitary evolution of vectors in Hilbert space is the basis for the formulation of quantum field theory (QFT). Wallace has strongly made the point that the MWI is the only viable interpretation of QFT as practiced by particle physicists and cosmologists currently on the table (\cite{wallace_plurality_2020}).

\subsection{Time in quantum mechanics}\label{sec:Time}

One area of investigation in quantum foundations, which is not often mentioned in discussions of the measurement problem, is the nature of time itself. In textbook formulations of the theory, time enters as a background parameter in postulate \textbf{P2}. This is in contrast to spatial coordinates, which enter as dynamical variables represented by Hermitian operators. The incompatibility between the Newtonian concept of time in quantum mechanics and the dynamical time of general relativity is referred to as the `problem of time' (\cite{anderson2012problem}).

With this in mind, recent years have seen much work devoted to the development of an alternative treatment of time in quantum mechanics. In particular, several works (\cite{giovannetti_quantum_2015,maccone_quantum_2020,hohn2021trinity}) develop a dynamical and relational notion of time, going back to an original suggestion by \cite{page_evolution_1983} in which quantum dynamics emerge from the correlations between spatially distinct subsystems of the universe. This methodology can be characterized as an `all-at-once' approach (\cite{adlam2022laws}), since it starts with the imposition of global constraints in the form of one (or more) atemporal Wheeler-DeWitt equations satisfied by the universal wavefunction (\cite{dewitt1967quantum}). 

Another alternative to the standard approach to quantum time is to introduce more than one temporal degree of freedom. An example of this is seen in the two-state vector formalism (TSVF) (\cite{aharonov_time_1964,aharonov1991complete,aharonov_two-state_2008,cohen2017quantum}), which was generalized to the multiple-time approach to quantum theory (\cite{vaidman_problem_1987,aharonov_multiple-time_2009}). The multiple-time approach has been demonstrated to be equivalent to the process matrix formulation in \cite{silva_connecting_2017}, and to other multiple time formalisms in \cite{nowakowski2018entangled}. It was also recently extended by \cite{aharonov_each_2014} to the idea of treating each instant of time as a `new universe'. One can also view the treatment of each time point in a quantum measurement as a label on a distinct subspace of a larger time-extended system. In the `each time a new universe' (ETNU) philosophy both chronological and anti-chronological influences propagate along extended `bricks' of time. This idea was developed further and generalized so that all times are treated equivalently in the recently-developed fixed point formulation (FPF) of \cite{ridley2023quantum}, which implements a fully time-symmetric model within an `all-at-once' approach. Within the FPF, the ETNU idea is implemented at each time and each time orientation on the Keldysh time contour (\cite{keldysh_diagram_1964}). Events in time are modelled as `sources' and `sinks' for regions of the wavefunction in their future and past on this contour. The FPF was recently shown to resolve some of the causal loop paradoxes which plague dynamical retrocausal models of quantum mechanics, and to resolve conceptual issues with the TSVF (\cite{ridley2025time}). As part of a small (but growing) literature on retrocausal `all-at-once' models (\cite{wharton2015universe,wharton2020colloquium,adlam2022two,maghsoudi2024evolutionary}), the FPF takes a step in the direction of \emph{merging} the two alternatives to Newtonian time described above. 

The novel structures involved in the FPF and the multiple-time formalism do not introduce new ontological types, nor do they alter the core unitary structure of quantum dynamics - they just apply the composition postulate \textbf{P3} directly to the times involved in a quantum measurement. They provide promising new avenues for arguments relating the structure of the core ontology of quantum theory to quantum probability.

\subsection{Overview of this work}

This paper provides an overview of the puzzle of quantum probability in the MWI, and proposes a novel solution to it. Section \ref{sec:probability} discusses the inherent conceptual problems with the concept of probability itself, with particular focus on equivocation norms. Borrowing from traditional concepts of physical probability, I argue that a particular type of equivocation is suitable for the assignment of quantum probabilities. In Section \ref{sec:probability_MWI}, I provide an overview of proposed solutions to the incoherence problem, and argue that the concept of overlapping worlds within the consistent histories approach supplies the most useful solution to this problem. I also summarize and evaluate the most successful attempts to resolve the quantitative problem to date. In Section \ref{sec:FPF}, I argue that it is possible to ground a rule with the same operational content as \textbf{P4} in terms of a time-symmetric framework which is operationally equivalent to \textbf{P1-P3}, even if \textbf{P4} in the form stated is not included as a postulate in the theory. That is to say, the predictive content of quantum mechanics is all present in an objective physical structure - the universal wavefunction - and the wavefunction contains structures over which we should directly distribute our degrees of belief.

\section{What is probability?}\label{sec:probability}

In discussions of probability in the MWI, it is often presumed that the notion of probability itself is uncontroversial, and has somehow been sorted out by philosophers and mathematicians. But, as \cite{papineau_fair_2010} emphasizes, there are real problems in the foundations of probability which make no reference to quantum mechanics. These problems often infiltrate and undermine discussions of probability in quantum mechanics. So before discussing solutions to the problems with probability in the MWI, it is useful to clarify which notion of probability is fit for this purpose.

\subsection{The epistemic/physical distinction}

The ontology of probability comes in one of two kinds: in \emph{epistemic} theories, probabilities are subjective degrees of belief, whereas in \emph{physical} theories probabilities are conceived as objective chances (\cite{carnap_two_1945,hacking_emergence_2006}). 

In the epistemic case, the ontology of probabilities is strictly mind-dependent, and so the concept of the probability of a proposition is often cashed out in terms of the betting odds that an agent would be willing to accept in a fair bet on the truth of that proposition (\cite{ramsey1926truth}). Rules which make numerical assignments of epistemic probabilities are \emph{normative} in nature - they prescribe the degrees of belief that a rational agent \emph{ought} to have about some proposition. This is the basic position of the various Bayesian interpretations of probability, which then go on to disagree heavily about which rules for assigning degrees of belief are rationally required (\cite{williamson_defence_2010,eagle_philosophy_2010}). 

In the case of physical objective chances, the theory of probability is \emph{descriptive} in nature, referring to facts about the structure of the world. Physical theories of probability include various frequentist interpretations going back to \cite{mises_probability_1939}, which identify relative frequencies accumulated over sequences of measurements with the objective chances. They also include the idea of probability as a physical `propensity' or intrinsic chancy disposition of the world (\cite{popper_propensity_1959,giere_objective_1973}). In propensity theories, frequencies may be used to confirm an assignment of objective chance, but they cannot be identified with it. Instead, the chances are conceived in opposition to determinism, as unavoidably random features of reality. 

In order to make good predictions about quantum systems, we therefore need rules connecting epistemic and physical states, i.e. which must identify facts about the state of the world that inform the assignment of rational degrees of belief. Following the terminology of \cite{strevens_objective_1999}, I refer to such rules as \textit{coordination principles}. One famous example of a coordination principle is the Principal Principle of \cite{lewis_subjectivists_1980}: 

\textbf{Principal Principle} - degrees of belief of a rational agent should be set equal to the objective chances.

According to Lewis, the chance is a function of time, and has no dependence upon the language or evidence of any agent. The existence of the objective chance function is a metaphysical thesis which accounts for the rationality of any action, expressed in terms of rationally assigned betting rates. This is an attractive thesis compared to other physical theories of probability. For instance, frequentists provide no incentive to the gambler who wishes to maximize his pay-off in the short-term. Rather, betting in accordance with objective chance in the long-run is a special application of the principle that one should always bet in accordance with the objective chance.

\subsection{Equivocation norms}\label{sec:equivocation_norms}

It is common practice, especially amongst quantum information theorists, to take the above as a complete summary of the substantial differences in the meaning of probability. The duality between subjective and objective probabilities has lead directly to the `Janus-faced' ontology of the wavefunction discussed by modern no-go theorems (\cite{pusey_reality_2012,colbeck_is_2012,leifer_is_2014}), which posit an underlying physical `ontic state' $\lambda$ uniquely determining the wavefunction. The quantum theory is then called $\psi$-ontic if the wavefunction can be shown to uniquely determine the ontic state, and $\psi$-epistemic if distinct quantum states are compatible with the same ontic state (\cite{spekkens_evidence_2007,harrigan_einstein_2010}). The conceptual debt of this approach to the distinction between epistemic and physical probabilities in the philosophy of probability is clear. 

But the real roots of the modern theory of probability lie in the classical theory, developed in the years 1650-1800 by the likes of Cardano, Pascal, Bernouilli, Leibniz, and culminating in the work of Laplace, whose probability theory Feynman was referring to in contradistinction to quantum probabilities (\cite{hacking_emergence_2006}). In the papers published by these early modern thinkers there is a recurring tendency to associate probability with notions of weighted possibility. \cite{hacking_equipossibility_1971} points out that the concept of possibility used in these theories also had a dualistic character - it sometimes referred to an epistemic notion, at other times to possibility as the `ease' with which physical states came into being. Yet the distinction between these different types of possibility is usually forgotten and the classical theory is characterized purely in terms of the following rule, due to \cite{laplace_theorie_1812}: 

\textbf{Principle of Indifference} - assign all possibilities an equal degree of belief when one has insufficient reason to do otherwise.

This is an example of an \emph{equivocation norm} - it tells a rational agent to equivocate their degrees of belief across a set of possible, but in some sense equally realizable outcomes in an experiment (\cite{williamson_defence_2010}). It can be motivated by a type of epistemic conservatism: one should not privilege any possibility over any other unless one has evidence to the contrary - spreading out your degrees of belief symmetrically automatically eliminates unjustified bias. 

However, there is a huge difference between assigning equal degrees of possibility on the basis of ignorance, and choosing your equipossible outcomes on the basis of positive information about a physical symmetry intrinsic to a system whose outcomes are unpredictable. In the latter case, there is much more going on than counting and combinatorics. A tossed die is a complex physical system which can land in one of a set of six physically equivalent macroscopic states. These states are physically equivalent because the set of individual outcomes corresponding to a small continuous range of initial conditions will contain each of the six sides face-up an equal number of times. This set of circumstances was called \textit{microconstancy} by \cite{strevens_objective_1999}. 

For the purposes of this work, I merely note that in a physical system, `equal possibilities' are directly explained by physical symmetries. They are explained by facts about the world, not facts about our ignorance, and it is only when one possesses such relevant facts about a physical system that one can equivocate over its possible states. Moreover, when such facts are available, we can replace the \textbf{principle of indifference} with a more empirically-based coordination principle:

\textbf{Principle of calibrated indifference} - assign equivocal degrees of belief over the physical possibilities in a dynamical system when the system contains a relevant structural symmetry with respect to these possibilities.

Setting aside for now, exactly what structural symmetries are `relevant', this type of coordination principle is quite different in nature to the Principal Principle. Whereas the latter seems to involve a metaphysical commitment to objective chance, there is no such commitment here. There is instead a commitment to distributing degrees of belief in accordance with the symmetries of the underlying physical system, which are present regardless of whether or not the system is observed. On this account, we should commit to the following type of inference:

\begin{align}\label{eq:implication}
\mathrm{\boldsymbol{\mathbf{\textrm{Physical Structure}}}}\longrightarrow\boldsymbol{\mathbf{\textrm{Rational Degrees of Belief}}}
\end{align}

By `Physical Structure', I refer to lawlike structural types which find a comfortable place in the ontology of our physical theories, not to instances of structure in the spatiotemporal organization of material objects. The inference (\ref{eq:implication}) is the fundamental implication which underlies any statistical scientific model, bar none. We should be suspicious of probability assignments, especially those invoked in physics, which reverse the direction of this implication.   

The source of an observer's uncertainty may simply lie in a combination of extreme initial condition sensitivity and uncontrollable initial conditions, as in the case of a classical coin toss (\cite{strevens_objective_1999}). For instance, objective probabilities in the dBB theory are attributed to uncertainty in the initial conditions of the system, and are therefore much closer in character to the microconstant view of classical statistical mechanics, as measures placed over trajectories corresponding to different initial conditions. In collapse-based interpretations of quantum mechanics, there appears to be a distinctly quantum type of objective chance, in which complete control of a system does not yield predictivity of a single-outcome measurement (\cite{brown_everettian_2020}). I discuss the type of probabilities which can appear in the MWI in the next section, but for now simply note that they cannot be the objective chances of physical collapse theories or the initial-condition uncertainties imposed by dBB. 

It is sometimes argued that the implication from structure to degrees of belief ought to be augmented by an additional calibration norm. This is the position of Objective Bayesianism, which argues that, even if physical structure underdetermines the rational degrees of belief, i.e. if a randomization process is not sufficient to justify a unique probability assignment, then we should be maximally equivocal over the possibilities (\cite{williamson_defence_2010}). This is formalized in the Maximum Entropy Principle (MEP), which instructs rational agents to assign those probabilities to a set of possibilities which maximize the Shannon entropy, subject to constraints. This approach to probability assignment was championed by \cite{jaynes_information_1957}, who argued that it sufficed for an explanation of the probabilities in statistical mechanics. However, equivocation without knowledge of the relevant physical symmetries can be a dangerous business. Indeed, there exists an entire class of `Bertrand-style' paradoxes (\cite{bertrand_calcul_1907}) in which applications of the uniform Lebesgue measure imply assignments of different probabilities to the same regions of an outcome space when the outcome space is re-parameterized. Jaynes devised extensions of the MEP which imposed further symmetry conditions on the probability measure (\cite{jaynes_well-posed_1973}) to tackle the Bertrand-style paradoxes. But he turned out to be ultimately unsuccessful in assigning a unique measure with this strategy in all cases - see (\cite{van_fraassen_laws_1989}) for a full account. Thus it seems one should avoid, as van Fraassen puts it, 

\begin{quote}
\emph{"the siren melody of empirical probabilities determined a priori on the basis of pure symmetry considerations"}
\end{quote}

Instead, an observer should practice indifference only if it is calibrated by the fundamental physical structure they are carrying out observations with respect to. Recent work on Maximum Entropy approaches to Bertrand-style paradoxes emphasizes that the accurate implementation of physical constraints is the missing ingredient for the completion of the Objective Bayesian program (\cite{parker2023maximum}). On this account, rational observers should choose to distribute their beliefs with respect to physical aspects of a system which are \emph{equally possible} for that observer, in the sense that the observer does not know in which part of the system they (or their measurement apparatus) are located. These possibilities could be construed to make up the total region of reality which is consistent with the experience of the observer, in the sense of being outcomes that their best physical theory allows them to consistently reach from their present state. If this is the source of their uncertainty, then the quantity to distribute our beliefs over, or to normalize probability assignments by, is surely the total region of accessible reality itself, where `accessibility' here refers to consistency with any physical constraints imposed on that region. Within the MWI, in accordance with \textbf{wavefunction monism}, reality is represented by $\left|\Psi_{U}\right\rangle$  directly. We can envisage cutting reality up into equally-sized pieces, or \textbf{units of wavefunction}. Different branches of the universal wavefunction may contain more or less of this basic substance. Reality has different-sized parts; all branches occur, but some occur more than others. Thus, the \textbf{principle of calibrated indifference} in this context is indifference with respect to \textbf{units of wavefunction}. We further refine this principle in the next Section with a view to producing an operational concept of probability in the many worlds setting.

\section{What is probability in many worlds?}\label{sec:probability_MWI}

In the Introduction, as in most discussions of probability in many worlds approaches to quantum mechanics, two major problems were posed: the \textbf{incoherence problem} and the \textbf{quantitative problem}. In this section I discuss solutions to both.  

There is a broad division between two kinds of Everettian strategies for answering the latter question: 

\textbf{Probability-first} - the concept of probability directs us to modify the formal rules of quantum mechanics, and add whatever structure is needed to account for the needs of the best theory of probability.

\textbf{Structure-first} - the core theory is all contained in the ontological and dynamical postulates and these imply the structure which calibrates degrees of belief.

In other words, we can either assume some notion of probability and try to extract the Born rule from this notion by applying a criterion for rational degrees of belief in the quantum context, or we can take the quantum theory as primitive and locate structures described by the Born measure within that theory. 

On the \textbf{probability-first} approach, the Born rule is an explication of some pre-theoretical notion of probability which is imposed from outside the core theory. This has both formal and ontological implications for quantum mechanics - for instance, if you are a propensity theorist then you may look for some concept of real objective chance, or something which can `play the role' of chances in the MWI (\cite{saunders_many_2010,papineau_fair_2010}).  

In \textbf{structure-first} approaches, the Born rule is not an explication of probability, but an explication of that part of the core quantum theory which looks like the Born rule. That is to say, structures with the form of Eq. \eqref{eq:Born} have a definite place in the ontology of the theory, which is wavefunction-based, and are explicated in the theoretical terminology of the core theory. If the core theory does not contain some structure which is the Born measure, one might decide to simply add the Born measure by hand, under some reasonable interpretation of its meaning (\cite{vaidman_derivations_2020}). But we should also reserve the option to \emph{modify the core theory} so that this structure appears automatically in descriptions of measurement processes. If, moreover, this can be done without modifying the \emph{character} of the core theory, so that it is still a theory about states in Hilbert space evolving unitarily, one might start to trust the new formulation.  

\subsection{The incoherence problem}

\subsubsection{Stochastic illusions}

The MWI is a deterministic interpretation of quantum theory, and therefore it rules out objective physical chances resulting from random dynamics. However, observers in a quantum measurement may still have uncertainty about the future - somebody who carries out a photon detection experiment does not know where a given photon will land, although it will land somewhere. The problem of reconciling uncertainty about the future with deterministic unitary dynamics of the universal wavefunction is the \textbf{incoherence problem}, and it has been viewed as a particularly pernicious problem for the MWI (\cite{dawid2015many,barrett_typical_2017}). 

The incoherence problem is not just the problem of accounting for probability within a deterministic theory. Given an initial condition distribution, the dBB theory is also deterministic with respect to time evolution, although the initial state of the system represents an epistemic constraint. In the MWI, there is no uncertainty in initial conditions or in the subsequent dynamics of the system, so it is very difficult to account for any recognizable notion of probability. It is in this sense that probability itself was called an `illusion' by Vaidman in the context of the MWI - there is no account of physical probability available that can be directly imported from other branches of science to describe the probabilities of worlds in $\left|\Psi_{U}\right\rangle$  (\cite{vaidman_schizophrenic_1998,vaidman_many-worlds_2021,vaidman_probability_2012,groisman_measure_2013,vaidman_quantum_2014,vaidman_derivations_2020,vaidman2022many}). The quantum state is completely determined at all times in the past, present and future. Therefore, Vaidman's idea can actually be separated into two illusions:

\begin{enumerate}
    \item The \textbf{illusion of uncertain preparation} - there is static indeterminism in the specification of the wavefunction at any time.
    \item The \textbf{illusion of chance} - there can be dynamical indeterminism in the temporal evolution of the wavefunction.
\end{enumerate}

In other words, since neither static nor dynamical indeterminism is present in the postulates \textbf{P1-P3}, they cannot explain probability in quantum mechanics. The dBB theory falls victim to the \textbf{illusion of uncertain preparation} and physical collapse theories fall victim to the \textbf{illusion of chance}. In what follows, I attempt to rescue a concept of probability in the MWI which bears ancestral similarity to the classical approach.

\subsubsection{Self-locating uncertainty}\label{sec:self-locating}

As \cite{wilson_objective_2013} has pointed out, the incoherence problem rests on a `one model one world' assumption: truth about the future of the world must refer to all branches of the wavefunction taken together. But this is not so - a proposition can be \emph{indexical} - true in one world, or one set of world-branches, but not in the global collection of branches making up the complete wavefunction. Actuality itself, for branch-limited observers, is indexical - a proposition such as `this world is one of many' can only be true when truth is `centered' on a specific branch of the wavefunction, i.e. its truth depends on the location at which it is asserted (\cite{wilhelm2021centering,wilhelm2022centering,wilhelm2023centering}). Such self-locating propositions have the peculiar quality of being true by virtue of the centered nature of the proposition, but they also require that the referent of the proposition is objectively situated within a larger structure. So the truth value of the proposition depends upon where it is uttered, but also upon objective structural facts relating all such locations on a map of reality. 

It is possible to possess self-locating uncertainty - uncertainty about the truth of a self-locating proposition in which the source of uncertainty is the self-location itself. Several authors have claimed that it is uncertainty with respect to a `center' at which the proposition is true which can provide the source of uncertainty in the MWI (\cite{vaidman_schizophrenic_1998,vaidman_quantum_2014,wilson_everettian_2012,wilson_objective_2013,saunders_many_2010,sebens_self-locating_2018,mcqueen_defence_2019}). This type of probability is epistemic but objectively grounded - it derives from a lack of knowledge regarding location within the real objective branching structure of the wavefunction. 

More formally, if observer $O$ assigns finite self-locating probability $\textrm{Pr}\left(P\right)$ to some property $P$, then there is at least one world in the wavefunction in which $P$ is instantiated and $O$ is uncertain whether or not they exist in some such world. 

The truth of propositions (and therefore the accuracy of self-locating uncertainties) may also be time-indexical (\cite{builes_time-slice_2020}). The relativization of truth to time, and the possibility of distinct epistemic centers located at different times, supplies an important case of self-locating uncertainty. Answers to problems involving this type of self-locating uncertainty usually hinge on acceptance of an indexical version of the \textbf{principal of indifference} due to \cite{lewis_sleeping_2001}: 

\textbf{Center indifference} - given an uncentered possible state of affairs containing some number of epistemic centers, rational degrees of belief are distributed evenly across those centers.

In the case of temporal self-locating uncertainty, the enumeration of epistemic centers is tightly connected to the metaphysics of time itself (\cite{builes_self-locating_2019}). Specifically, on a \textbf{presentist} conception of time, the uncentered `stage' consists in three-dimensional space at a single instant - it is impossible to apply \textbf{center indifference} with respect to time on this view, although other types of self-locating uncertainty are possible. On an \textbf{eternalist} view of time, all times in the past, present and future are equally real and epistemic centers can be distributed anywhere in the four-dimensional space-time manifold. The latter picture is standard amongst cosmologists, and so Everettian worlds are sometimes replaced by talk of spacetime world-tubes, or branches of $\left|\Psi_{U}\right\rangle$  which are extended in time (\cite{wallace_worlds_2002}). We
also note here the existence of theoretical (\cite{taylor2004entanglement,adlam2018spooky}) and experimental (\cite{megidish2013entanglement,cotler2017experimental}) work demonstrating the feasibility of \emph{entanglement in time}, whereby it is demonstrated that particles which have no temporal coexistence share non-classical interdependence. 
Thus, I come down firmly on the side of eternalism. 

If it is to represent all of reality, and if we subscribe to the \textbf{eternalist} viewpoint such that reality includes all times, then the universal wavefunction must represent all times, and is therefore a time-extended object. To consider $\left|\Psi_{U}\right\rangle$ in its entirety requires us to move to an atemporal, block universe picture of reality (\cite{stoica2021post}). Probabilities must therefore be defined over time-extended regions of the wavefunction, which we can think of as broken down into very small time-extended \textbf{units of wavefunction} serving as the epistemic centers described above. The \textbf{principle of calibrated indifference} that was developed at the end of Section \ref{sec:equivocation_norms} is therefore upgraded to \textbf{center indifference} with respect to \textbf{units of wavefunction}.

\subsubsection{Overlapping vs divergent worlds}\label{sec:overdiverge}

\begin{figure}[htp]
\centering
  \begin{subfigure}{0.5\linewidth}
    \includegraphics[clip, width=\linewidth]{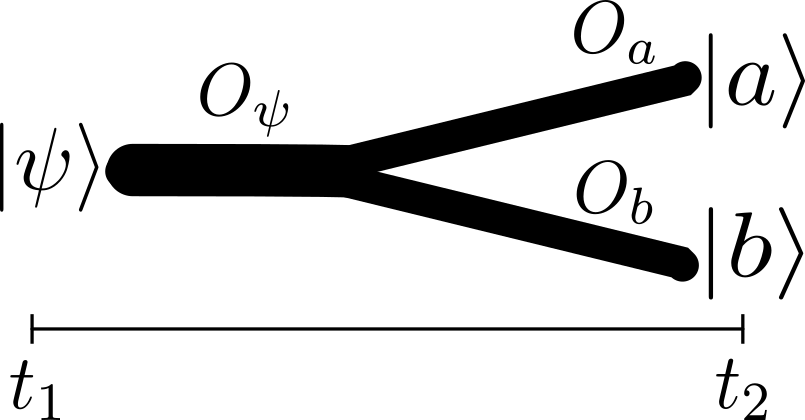}
    \subcaption{Overlapping worlds with new observers $O_{a}$ and $O_{b}$ coming into existence after branching.} \label{fig:1a}
  \end{subfigure}%
  
  \begin{subfigure}{0.5\linewidth}
    \includegraphics[clip, width=\linewidth]{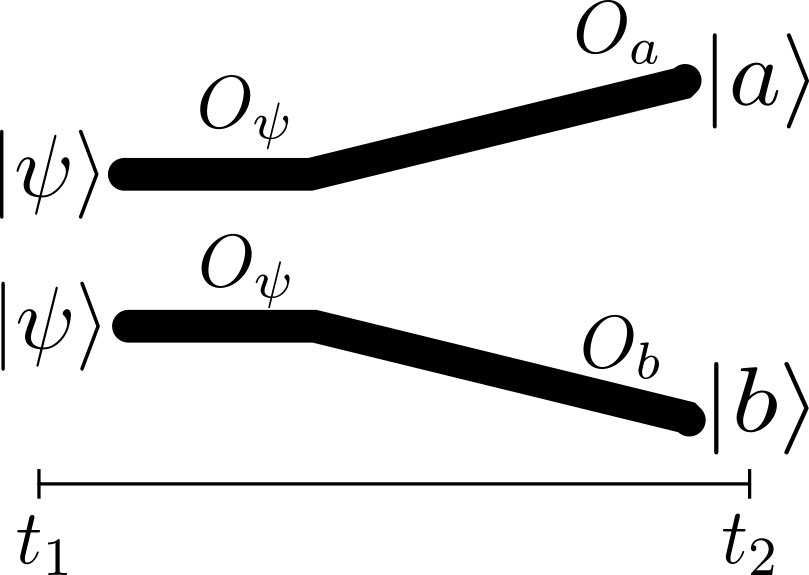}
    \subcaption{Diverging worlds with world-indexed time-extended observers having temporal parts $O_{\psi}$ at $t_{1}$ and $O_{a}$ at $t_{2}$ on the upper branch, and $O_{\psi}$ at $t_{1}$ and $O_{a}$ at $t_{2}$ on the lower branch.} \label{fig:1b}
  \end{subfigure}%
  \caption{}
  \label{fig:over_diverge}
\end{figure}

We now come to the peculiar time-dependent process of branching. Fig. \ref{fig:over_diverge} (a) represents the wavefunction prepared in state $\left|\psi\right\rangle$ at $t_{1}$ which splits into two worlds described by the states $\left|a\right\rangle$ and $\left|b\right\rangle$ before being measured at time $t_{2}$. Both worlds form parts of a single branching wavefunction. For an observer part of the initial state, $O_{\psi}$, whose part of the wavefunction also experiences this splitting process, a very unusual metaphysical scenario arises. For they seem to be located, more or less unaltered, in every branch! So how can it make sense to talk about their uncertainty of self-location? They seem to have temporal parts on the single pre-branching `trunk' of the wavefunction and on each of the post-branching descendant worlds. Moreover, if we try to relativize identity to branches, obtaining two post-branching observers $O_{a}$, $O_{b}$, it appears that these two individuals share the temporal part $O_{\psi}$ prior to the branching. They are called \emph{overlapping} persons - identities which at certain times are numerically identical with one another, and which have identical memories of those times when they overlap. Such `transworld identities' make it very difficult to maintain ordinary beliefs about the future - it is hard to account for our everyday single-world experience if we are continually undergoing splitting in this sense.  

One response to the problem of future beliefs is to restrict situations of genuine self-locating uncertainty to post-measurement pre-observation uncertainty. This is the position advocated by Vaidman, who argues that the splitting of the universe is a physical event that fundamentally alters the epistemic situation for an observer in the universe (\cite{vaidman_probability_2012}). On Vaidman's account, assignments of self-locating uncertainty can only be made at certain times - a rational agent should have no uncertainty about the future, but if they know that other branches exist parallel to their own branch now, they may have epistemic uncertainty regarding which branch they are in. In this sense, assignments of rational degrees of belief are inherently tied to the objective branching structure of $\left|\Psi_{U}\right\rangle$. The \textbf{illusion of chance} comes about by a physical process in which new `centers of subjectivity', all with identical memories of the past, come into existence every time there is a genuine branching in the wavefunction (\cite{albert_probability_2010,vaidman_probability_2012,groisman_measure_2013,vaidman_quantum_2014}). The real source of uncertainty lies in the multiplicity of observers post-measurement, which did not exist prior to the measurement being carried out. This is a case of branch-indexicalism coupled to time-indexicalism with regard to identity. The two post-branching observers $O_{a}$ and $O_{b}$ in Fig. \ref{fig:over_diverge} (a) are numerically distinct from $O_{\psi}$, but share enough properties over time to function as the same observer in each world-branch. 

This notion of pre-observation post-measurement self-locating uncertainty was criticized as coming `too late in the game' by \cite{albert_probability_2010}. Even if identity is in some sense intrinsically time-indexical, the actual experience of an observer is that of continuity between their pre- and post-measurement selves. Moreover, there is a practical objection - probabilities in quantum mechanics are indeed used to predict the future - it is natural to be uncertain about what will be observed in a single-case experiment.  

Perhaps most problematically on this account, a large burden is placed on the sudden emergence of new personal identities when a branching process occurs (\cite{cunningham_branches_2014,marchildon_multiplicity_2015}). If the probabilities of quantum mechanics depend upon the existence of entities like this, and can only be expressed from the perspective of such entities, then the MWI veers dangerously close to a subjectivist privileged perspective account of statistical prediction, which is precisely the kind of account that it was constructed to avoid. On this understanding, it becomes very difficult to avoid the conclusion of \cite{albert_probability_2010}: 

\begin{quote}
    \emph{“those uncertainties have nothing whatever to do with objective metaphysical features of the world”.}
\end{quote} 

In the parlance of Section \ref{sec:equivocation_norms}, it becomes much more difficult to identify the objective physical structure which grounds rational belief assignment, and therefore much more difficult to uphold the implication in Eq. \eqref{eq:implication}. If it is maintained on this picture that the new epistemic centers which appear during the measurement are metaphysically primitive, then we find ourselves in the strange situation whereby an agent can freely choose to split the world (and themselves) into numerically distinct, but similar entities, such that the time they choose to do the measurement becomes ontologically significant for the entire universe. For such an agent, this seems like a privileged point of view indeed!

Fortunately, \cite{lewis1976survival} identified an alternative notion of branching identity which makes more sense of ordinary belief assignments about the future. Instead of a single branching process, we can understand the process in Fig. \ref{fig:over_diverge} (a) as a set of two processes occurring in parallel in mutually exclusive worlds, as illustrated in Fig. \ref{fig:over_diverge} (b). In this case the identity of the observers $O_{a}$, $O_{b}$, is said to be \emph{diverging} - they may have been indistinguishable prior to the measurement, but they were numerically distinct. If they possess identical physical properties at times prior to the measurement time, this means that instantiations of those properties are of identical \emph{type}, not identical \emph{tokens} of the same type of property - all their properties were identical except for that of branch location (\cite{wilson_everettian_2012}). 

The formalism of quantum mechanics itself does not distinguish between the diverging and overlapping pictures, although the diverging picture is closer in spirit to approaches such as the path integral formalism or the consistent histories approach (discussed in Section \ref{sec:CH}). But recent years have seen authors such as Wallace, Saunders and Wilson all come down on the side of the divergence model on the basis of purely metaphysical considerations (\cite{saunders_branching_2008,wallace_emergent_2012,wilson_objective_2013}).

First of all, it is easy to see how divergence makes sense of our ordinary beliefs about the future. In Fig. \ref{fig:over_diverge} (b) it is illustrated how each observer is always located in a single world on a single time-extended branch of the wavefunction. They have genuine uncertainty about what they will observe (and which observer they are), and so when they perform the measurement they record one of the outcomes $\left|a\right\rangle$, $\left|b\right\rangle$  which resolves their self-locating uncertainty. \cite{hedden_time-slice_2015} has argued that any involvement of a theory of personal identity over time in an account of truth-assignments is problematic - in the many worlds framework, assignments of self-locating probability by an observer in one of the worlds seem to rely on that observer surviving the measurement itself. But why should the predictions of quantum mechanics have anything to do with personal identity whatsoever? Why does an electron care about my survival? The answer, of course, is that it does not. \cite{lewis2007uncertainty} strongly makes this point in his discussion of the branching process - the deep structure of nature does not distinguish between accounts of personal identity over time. An account of personal identity, no matter how conceptually satisfying, has no bearing on the deeper mechanics which should explain a fundamental physical probabilistic law. The Born rule is just this kind of law. 

I therefore follow \cite{saunders_branching_2008}, and shift the source of self-locating uncertainty, the loci of our epistemic centers, from observers in worlds to worlds themselves, which are just regions of the wavefunction. On this understanding, it is the real physical structure of many parallel diverging world-tubes, which may be indistinguishable at certain times, which justifies the pre-measurement uncertainty of an observer. 

In addition, the diverging worlds picture contains no intrinsic metaphysical difference between past and future - worlds are not created at the whim of an experimentalist, and observers within each parallel world exist in an intuitive single-world manner, from cradle-to-grave (\cite{deutsch_quantum_1985}). That they can do so also depends on the emergence of quasiclassical world branches which admit stable orthogonal macrosopic states over time. \cite{wallace_emergent_2012} has argued extensively for the emergence of such structures via decoherence as an automatic outcome of the axioms of quantum mechanics. Moreover, recent numerical work by \cite{strasberg2024} confirms the picture of classical worlds emerging from the quantum state for a large class of model Hamiltonians. 

We now consider the observer in the branching structure of Fig. \ref{fig:over_diverge}. This observer is a macroscopic object living in a world governed, for all practical purposes, by the familiar laws of classical mechanics, chemistry and biology. The structures which this observer finds useful are \emph{emergent} from the underlying microscopic physics, enabling them to understand their experience with the minimum number of coarse-grained degrees of freedom. These structures are objective consequences of the underlying physics - they typically arise from the spatial locality of particle interactions and the rapid destruction of interference between macroscopic degrees of freedom via decoherence (\cite{riedel_objective_2016,strasberg2023shearing}). The objects of experience may therefore be instantiated in the wavefunction as trajectories through an appropriately coarse-grained property space, indexed by time. 

\subsubsection{Decoherent histories}\label{sec:CH}

Trajectories through property space is precisely the picture described by the decoherent histories formulation of quantum theory (\cite{gell_classical_1993,gell1994time,gell_quantum_1996,saunders1995time}), which offers a natural extension of the MWI to entire temporal regions (\cite{wallace_emergent_2012}). In this formulation, the `observer' as distinct from the system is not required. What is required is a collection of sequences of physical properties, or histories, which are mutually exclusive and run parallel to each other in time. Observers located in different histories tell different stories about the physical properties of their world - there are times at which they do not agree on the value of all physical observables. To be precise, in the consistent histories formulation of quantum mechanics, single time quantum states are replaced with sequences of states 
\begin{align}
    \mathbf{\alpha}\equiv\left[\alpha_{N},\ldots,\alpha_{1}\right]
\end{align}
where $\alpha_{i}$ denotes a configuration of the system at time $t_{i}$ in some coarse-grained, time-localized Hilbert space $\mathcal{H}_{t_{i}}$, and there is some time ordering $t_{0}<\ldots<t_{N}$. Given the initial preparation in state $\left|\psi_{1}\right\rangle$  at time $t_{1}$, the probability to measure $\alpha_{2}$ at the subsequent time $t_{2}$ is given by the Born rule
\begin{align}\label{eq:probability}
    p\left(\alpha_{2}\left(t_{2}\right);\psi_{1}\left(t_{1}\right)\right)=\left|\left\langle \alpha_{2}\right|U\left(t_{2},t_{1}\right)\left|\psi_{1}\right\rangle \right|^{2}
\end{align}
where we introduce the unitary time propagator implied by the time-dependent Schr{\"o}dinger equation in Eq. \eqref{eq:TDSE}, with chronological time-ordering imposed by the operator $\hat{\textrm{T}}$:
\begin{align}\label{eq:propagator}
    U\left(t_{b},t_{a}\right)=\hat{\textrm{T}}\left[\exp\left[-\frac{i}{\hbar}\int_{t_{a}}^{t_{b}}d\tau \hat{H}\left(\tau\right)\right]\right]
\end{align}
the propagator provides a mapping between local-time Hilbert spaces, $U\left(t_{b},t_{a}\right):\mathcal{H}_{t_{a}}\rightarrow\mathcal{H}_{t_{b}}$ spanned by the appropriate coarse-grained bases. 

The expression \eqref{eq:probability} can also be written in terms of a projection onto $\alpha_{2}$ in the Heisenberg representation
\begin{align}
    P_{\alpha_{k}}\left(t_{1}\right)=U^{\dagger}\left(t_{k},t_{0}\right)\left|\alpha_{k}\right\rangle \left\langle \alpha_{k}\right|U\left(t_{k},t_{0}\right),
\end{align}
with initial density operator $\rho_{1}=\left|\psi_{1}\right\rangle \left\langle \psi_{1}\right|$:
\begin{align}\label{eq:alpha_sequence}
    p\left(\alpha_{2}\left(t_{2}\right);\psi_{1}\left(t_{1}\right)\right)=\textrm{Tr\ensuremath{\left[P_{\alpha_{2}}\left(t_{2}\right)\rho_{1}P_{\alpha_{2}}\left(t_{2}\right)\right]}}
\end{align}
This can be generalized so that the history sequence in Eq. \eqref{eq:alpha_sequence} is represented as an ordered sequence of projections, giving the probability of a history:
\begin{align}
    p\left(\mathbf{\alpha};\psi_{1}\left(t_{1}\right)\right)=\textrm{Tr\ensuremath{\left[P_{\alpha_{N}}\left(t_{N}\right)\ldots P_{\alpha_{2}}\left(t_{2}\right)\rho_{1}P_{\alpha_{2}}\left(t_{2}\right)\ldots P_{\alpha_{N}}\left(t_{N}\right)\right]}}
\end{align}
It is common to re-express this in terms of the so-called history operator:
\begin{align}\label{eq:history}
    \mathbf{C}_{\mathbf{\alpha}}\equiv\left[P_{\alpha_{N}}\left(t_{N}\right),\ldots,P_{\alpha_{2}}\left(t_{2}\right)\right]
\end{align}
with respect to which `record states' are defined (\cite{gell_classical_1993,wallace_emergent_2012}), providing a wavefunction-based representation of the histories of entire wavefunction branches
\begin{align}\label{eq:record}
    \left|\alpha\right\rangle \equiv\mathbf{C}_{\mathbf{\alpha}}\left|\psi_{1}\right\rangle 
\end{align}
and the probabilistic weights of histories are reformulated in terms of the decoherence functional:
\begin{align}
    p\left(\mathbf{\alpha};\psi_{1}\left(t_{1}\right)\right)=\mathbf{D}\left(\mathbf{\alpha},\mathbf{\alpha}\right)\equiv\left\langle \psi_{1}\right|\mathbf{C}_{\mathbf{\alpha}}^{\dagger}\mathbf{C}_{\mathbf{\alpha}}\left|\psi_{1}\right\rangle
\end{align}
The record states in Eq. \eqref{eq:record} exist in a tensor product history space of N time-local Hilbert spaces, each of which is a copy of the physical Hilbert space (\cite{isham1995continuous})
\begin{align}\label{eq:history_space}
    \mathcal{H}_{H}\equiv\mathcal{H}_{t_{N}}\otimes\ldots\otimes\mathcal{H}_{t_{1}}
\end{align}
i.e. the composition postulate \textbf{P3} is applied to instants of time themselves. $\mathcal{H}_{H}$ is said to be a \textbf{decoherent history space} iff for all $\alpha,\beta\in\mathcal{H}_{H}$ $\mathbf{D}\left(\mathbf{\alpha},\mathbf{\beta}\right)=0$ iff $\alpha\neq\beta$, the condition for decoherence between any pair of histories in the space (\cite{wallace_emergent_2012,riedel_objective_2016}). Each decoherent history in the space evolves in parallel with every other, establishing the simultaneous existence of non-overlapping quasi-classical branches. The universal wavefunction $\left|\Psi_{U}\right\rangle$ is then some linear combination of the $\left|\alpha\right\rangle$; a time-extended object.

The objective existence of each history, on a bird's-eye view of the wavefunction, is sufficient grounds for making an assignment of self-locating uncertainty for an observer localized to one history (\cite{wallace_emergent_2012}). A macroscopic individual can commit to the proposition that they are in one trajectory, because they exist at many points in time but cannot possess mutually exclusive values of their physical properties at the same time. This is not a hidden variables theory - the observer cannot predict the outcome of a measurement in the future of their branch. However, their branch future is determinate in the sense that, on an atemporal, bird's-eye view of the universal wavefunction, it is fixed in place. In this sense, it has the same metaphysical status as their past and so the consistent histories framework fits naturally into the \textbf{eternalist} view of spacetime as a four-dimensional manifold. Indeed, if unification is the ultimate goal, it must.

In the strict interpretation implied by \textbf{wavefunction monism}, the multiple-time structure $\left|\Psi_{U}\right\rangle$ is identified with a branching universe. An observer located in a single branch can extract all the features of their experience from it. The macroscopic objects of their experience are patterns which supervene on this structure, and their stability in time derives from the stability of this structure in time. So one should look to this structure for an explanation of the probability measure defined over a history. In his discussion of emergent branching structures, \cite{wallace_decoherence_2010} appears to make this connection:

\textit{“The existence of this branching is a robust (albeit emergent) feature of reality; so is the mod-square amplitude for any macroscopically described history.”}

Given this notion of emergent mod-square amplitudes, one could be forgiven for expecting such structures to require no further formal support or argumentation. Indeed, if we accept \textbf{wavefunction monism}, then any `feature of reality' is just a feature of the wavefunction: it must be visible in a bird's-eye map of the universe, and therefore in each multiple-time history state. So it is fair to ask the question: which emergent structure of the wavefunction is the mod-square amplitude? As \cite{lewis2009probability} correctly points out, there is no such structure, emergent or otherwise, in the standard MWI. 

If such structure cannot be seen directly in our formalism it may be that the mathematical formalism is lacking some structural and which therefore must be incorporated into the structural postulates of quantum theory from the outset. This strategy is \textit{logically minimal, physically maximal} - it grounds the mathematical theory of probability in physical ontology (\cite{ridley2023quantum}). 

\subsection{The quantitative problem} 

There have been many different attempted solutions to the \textbf{quantitative problem} of probability in the MWI. In his original paper, Everett thought that the Born probability measure could be attached to branches in a straightforward way via some additional constraints on probabilities (\cite{everett_relative_1957}). However, upon closer inspection, these constraints prove to be rather strong axioms which it is impossible to motivate from the quantum theory alone. They include, for instance, the requirement that the probability measure be a unique function of the modulus of the wavefunction, a condition which has no counterpart in classical probability theory.

Since Everett, there have been several attempts to derive the Born measure as a probabilistic measure over worlds. They typically involve the proposal of rationality constraints on degrees of belief in a universe described by the postulates \textbf{P1-P3} (\cite{deutsch_quantum_1999,wallace_emergent_2012,sebens_self-locating_2018}). Said strategies have come under increasing scrutiny (\cite{vaidman_derivations_2020}), as they seem to involve the derivation of a physical law from constraints on `rationality'. This is very suspicious. Why should nature care about what is deemed `rational'? The answer, of course, is that it does not. This type of argumentation reverses the implication in Eq. \eqref{eq:implication}, and therefore cannot answer the question that has been posed in this paper. 

Take the Deutsch-Wallace theorem (\cite{wallace_how_2010}). This is a decision-theoretical argument showing that, if an agent acts in accordance with a set of `rationality' axioms in an Everettian multiverse, then they will act as if the Born rule is true. Setting aside the validity of the derivation (\cite{mandolesi_quantum_2020}), the axiom choice is not compelling because (i) it is not clear that said axioms are all true or even consistent with one another - see \cite{mandolesi2019analysis} for a devastating critique - and (ii) the axioms are not motivated by structural physical facts.

I would like to ask how, exactly, the mod-squared amplitude measure is related to the postulates \textbf{P1-P3}, or to an equivalent formulation of ontology and dynamics. This is just the answer to the question posed in Section \ref{sec:equivocation_norms}, of how to derive rational degrees of belief from physical structure. It is also in keeping with Hartle's stipulation that one should prefer the derivation which reveals connections between elements of the theory.  

\subsubsection{Branch counting}\label{branch_counting}

As Everett's ideas were popularized in the 1970s, there were attempts to address the \textbf{quantitative problem} of the MWI through the \textbf{principle of indifference} applied to the number of wavefunction branches - branch counting arguments were made (\cite{dewitt_quantum_1970}) and discredited (\cite{kent_against_1990,araujo_probability_2019}). The idea of branch counting has recently seen some resurgence, with \cite{saunders_branch-counting_2021} proposing an alternative version of the branch-counting rule motivated by analogies with Boltzmannian statistical mechanics. 

Since the branching process is the cause of multiple observers possessing uncertainty in their self-location, the branch number might seem like the relevant physical quantity to equivocate over. But, as discussed in Section \ref{sec:equivocation_norms}, ignorance between a set of distinguishable possibilities has no bearing whatsoever on the way you should distribute your beliefs over them. Rather, one should look for a \textbf{principle of calibrated indifference} - what we are indifferent with respect to must have a specific kind of physical meaning. \cite{saunders_branch-counting_2021} appears to recognize this and so proposes a rule that sets branches (modelled as quasi-classical histories within the decoherent histories formalism) with equal amplitudes equal to the same probabilities. This principle is a step in the right direction, but still insufficient for the logically reductionist task at hand. It does not follow from either Saunders' frequentist conception of probability or from the core postulates of quantum mechanics.

The principle we seek must refer to what is in some sense equally physically realized, and in general wavefunction branches differ with respect to the proportion of the wavefunction they occupy. In the next section, I move from the notion that branches are equally realized to the idea that they come with different `thicknesses', which directly equate to this relative proportion.

\subsubsection{The measure of existence}\label{sec:measure_of_existence}

Since in the MWI everything which is realizable is realized on a different wavefunction branch, and together all branches make up the sum total of reality, I may state the \textbf{principle of calibrated indifference} as indifference between branches which occupy the same amount of reality. This is the basic intuition behind the simplest account of the Born measure in the context of the MWI: Vaidman's concept of the \emph{measure of existence}. According to Vaidman, to resolve the \textbf{quantitative problem} of the MWI we must add at least one mathematical postulate about the relative size of worlds. To define this measure, Vaidman makes a decomposition of the universe into the different worlds that are possible for the observer: 

\begin{align}
    \left|\Psi_{\textrm{U}}\right\rangle =\Sigma\alpha_{i}\left|\psi_{\textrm{world i}}\right\rangle 
\end{align}

Then the measure of existence of world $i$ is \emph{postulated} to have the value $\left|\alpha_{i}\right|^{2}$. On this account, the introduction of a probability law in the MWI is simpler than \textbf{P4}, as it omits collapse and links a physical property of the universal wavefunction directly to a rational belief assignment (\cite{tappenden_evidence_2011}):

\textbf{Born-Vaidman rule}

\textit{The rational degree of belief in an outcome of a quantum experiment is proportional to the total measure of existence of all worlds with that outcome.} 

The Born-Vaidman rule does not argue deductively from constraints on rational degrees of belief to theoretical predictions. \cite{vaidman_derivations_2020} instead makes the meta-argument that such constraints are an unnecessary complication because the logically parsimonious formulation is to postulate the mathematical form of the Born measure outright. Whereas this cannot be considered a \emph{derivation} of rational degrees of belief from the underlying sub-structure of reality, it at least does not reverse the implication in Eq. \eqref{eq:implication}, i.e. it gives a non-rigorous account of the physical property - the measure of existence - which characterizes rational belief assignments over wavefunction branches (\cite{groisman_measure_2013}). On this account, the objective nature of this sub-structure is confirmed directly by observed relative frequencies (\cite{mcqueen_defence_2019}). 

Scientific practice proceeds on the assumption that stable probability densities exist in nature, for instance in the decay rates of unstable isotopes (\cite{maudlin_what_2007,wallace_emergent_2012}). The confirmation of the Born measure from frequencies has been brought into question by \cite{albert_probability_2010}, due to the existence of “maverick” worlds in which Born rule statistics fail. But an experimentalist in a maverick world can always look to consistent natural phenomena in their world, such as the rates of nuclear decay or the fact that the sky is blue (\cite{mcqueen_defence_2019}), which are direct predictions of the Born rule. These features of the world confirm the Born rule independently of an individual run of measurements. 

The Born-Vaidman rule for the measure of existence, was supplemented by an argument in \cite{vaidman_probability_2012} and further developed in \cite{mcqueen_defence_2019}, which appealed to a physical symmetry principle - measurements performed on two physically identical spatial locations must yield the same results - and to the principle of local supervenience - whatever an observer experiences in a region of spacetime depends only on the physical description of that region. 

However, this symmetry argument rests upon the truth of empirical facts that are not contained in the postulates \textbf{P1-P3} of quantum mechanics (\cite{blackshaw2023probability}), but which have to be established separately. It is therefore not a logically clean axiomatic derivation. More problematically for this approach, in most cases quantum states with equal amplitudes do not correspond to spatially interchangeable outcomes, so this argument for the Born rule is not a general proof.

Moreover, a concept of an existence measure does not address the logical question of whether the mathematical probability law can be derived, or the physical question of which aspect of the wavefunction the existence measure can be identified with. Intuitively, a measure of existence is a measure of the amount of total reality possessed by a branch of the wavefunction. So first of all, this concept must be precisified. 

The Born measure should reflect a symmetry in the wavefunction which is precisely the equal amount of reality possessed by equally-sized regions of the wavefunction. If regions of the wavefunction have equal sizes, then they have equal probabilities, because the latter are set equal to the former. This is the strongest possible precisification of the Born-Vaidman rule, and it puts the onus firmly on \textbf{P1-P3} with no further rationality constraints involving the “siren melody” that van Fraassen warned of. 

We now turn to arguably the most `quantum' incarnation of this siren melody.

\subsubsection{Envariance}

There is a remarkable formal similarity between the leading strategies for a derivation of the Born rule in the MWI. As \cite{drezet2021making} has made clear, they all utilize a particular form of the \textbf{principle of indifference}, but instead of applying it to branches, it is applied to symmetries of composite states under operations performed on subsystems of the composite system. This particular symmetry, known as entanglement-assisted invariance or \emph{envariance}, was first made explicit in connection with derivations of the Born measure by \cite{zurek_environment-assisted_2003}, although Zurek does not completely subscribe to the MWI (\cite{zurek_quantum_2018,zurek2022quantum}). I therefore focus on the argument of Zurek.

Suppose one has a composite system made up of subsystems $A$ and $B$, represented by states in the Hilbert spaces $\mathcal{H}^{A}$ and $\mathcal{H}^{B}$. These subsystems may refer to a microscopic system and its environment, or to a system plus its observer, or to some partition into system, environment and observer, or, in decision-theoretical parlance, to the outcome and reward subspaces. 

We consider a generic entangled state  $\left|\Psi_{AB}\right\rangle \in\mathcal{H}^{A}\otimes\mathcal{H}^{B}$ in terms of orthonormal bases $\left\{ \left|a_{i}\right\rangle \right\}$  and $\left\{ \left|b_{i}\right\rangle \right\}$  spanning the Hilbert space of each subsystem, expressed as a Schmidt decomposition:

\begin{equation}\label{eq:AB_psi}
    \left|\Psi_{AB}\right\rangle =\overset{N}{\underset{k=1}{\sum}}\alpha_{k}\left|a_{k}\right\rangle \left|b_{k}\right\rangle 
\end{equation}

In Zurek's original formulation, A represents the system of interest, whereas B represents the environment of the system (\cite{zurek_environment-assisted_2003}). Envariance is precisely defined as follows:

$\left|\Psi_{AB}\right\rangle$  is envariant with respect to the transformation $U_{A}$ which acts solely on A if and only if there exists a corresponding transformation $U_{B}$ acting solely on the B subsystem which returns the transformed system to its initial state, i.e. when

\begin{equation}
    U_{B}\left(U_{A}\left|\Psi_{AB}\right\rangle \right)=\left|\Psi_{AB}\right\rangle,
\end{equation}

where $U_{A} \equiv \mathcal{U}_{A}\otimes\textbf{I}_{\mathcal{B}}$ and $U_{B} \equiv \textbf{I}_{A}\otimes\mathcal{U}_{\mathcal{B}}$ act only on the $A$ and $B$ subspaces, respectively. The argument of Zurek (and indeed of all symmetry-based derivations of the Born rule) begins with the consideration of a maximally entangled base case, in which the amplitudes of all terms in the superposition are equal

\begin{equation}\label{eq:AB-psi-entangled}
    \left|\overline{\Psi}_{AB}\right\rangle =C\overset{N}{\underset{k=1}{\sum}}e^{i\phi_{k}}\left|a_{k}\right\rangle \left|b_{k}\right\rangle
\end{equation}

Here, $C$ is a real-valued constant. Zurek goes on to argue in two steps, firstly that all states in a decomposition with the form in Eq. \eqref{eq:AB-psi-entangled} must have the same probability, and secondly that entangled states with unequal coefficients of the form in Eq. \eqref{eq:AB_psi} can always be reduced to the symmetric case of equal coefficients - the so-called `coarse-graining' step. 

The derivation of \cite{zurek_environment-assisted_2003} relies on several assumptions, termed `quantum facts', which do not assume the Born rule. These facts apply to the completely decohered state in Eq. \eqref{eq:AB_psi}, in which $A\equiv \mathcal{S}$ is assumed to represent the system of interest and $B\equiv \mathcal{E}$ represents the environment. In particular, it is assumed (i) the state of $\mathcal{S}$ alone uniquely determines the probabilities of measurement outcomes on $\mathcal{S}$ (ii) the state of the larger $\mathcal{S}+\mathcal{E}$ composite system determines the state of $\mathcal{S}$ (\cite{zurek_relative_2007}) and (iii) Laplace's \textbf{Principle of Indifference} applies to an observer in the environment considering system states that are equivalent up to a permutation. 

Setting aside the validity of the coarse-graining step, I object to the strategy chosen by Zurek in the equal-amplitude case. It is not clear what Zurek's probabilities \emph{are}, although they are assumed to be properties which supervene on quantum states. They serve to play the role of objective chances, and they are confirmed by frequencies, but where do they originate? What is the precise nature of the connection, on Zurek's view, between the ontology and the chances? Zurek recognizes that this is a problem and attempts to answer it thus (\cite{zurek2022quantum}):

\begin{quote}
    \emph{"We emphasize that in contrast to many other approaches to both classical and quantum
probability, our envariant derivation is based not on a subjective assessment of an observer,
but on an objective, experimentally verifiable symmetry of entangled states."}
\end{quote}

Thus, Zurek advocates for a kind of equivocation which is `backed-up' by the properties of entangled states. There is an element of objectivity in his argument - he uses quantum `facts' to formulate it. But these are facts about \emph{probabilities}, not states. They do not follow from postulates \textbf{P1-P3}. The only possible justification for these facts is that they are true for the relative frequencies observed on subsystems in nature - a topic of current experimental inquiry (\cite{deffner2017demonstration}).

Here, an Everettian should object that Zurek's probabilities would not be visible to a hypothetical observer with a bird's-eye view on nature, standing external to the wavefunction. From that vantage point, So although his derivation is \textbf{structure-first} in nature, it is not clear \emph{which} structures in the ontology are the probabilities. The probabilities are not included in the core theory. One could pursue Zurek's argument strategy in a much simpler way, and use the `quantum fact' that equal-amplitude states are equiprobable as the foundation for his derivation. One must, as \cite{ismael2009probability} has argued, specify what quantum probabilities are as part of the specification of the quantum theory itself.

In the next Section I build on the work done so far to develop a concept of quantum probability which would be apparent both to a bird's-eye observer who is party to the evolution of every wavefunction branch, \emph{and} to a branch-localized observer which uses this property of branches to calibrate their degrees of belief.

\section{The fixed-point formulation}\label{sec:FPF}

I have argued that the ubiquity of the Born measure in nature deserves an explanation which is deeply grounded in ontology and dynamics. In other words, a \textbf{structure-first} derivation of the Born rule is necessary. In order to tighten the connection between representational models and probabilities, I argued in Section \ref{sec:probability} for a \textbf{principle of calibrated indifference} which asserts that we should distribute our beliefs with respect to \textbf{units of wavefunction}. I extended this analysis in Section \ref{sec:probability_MWI}, arguing that the epistemic situation of an observer located somewhere in this wavefunction is one of self-locating uncertainty - they do not know in which bit of wavefunction a measurement will turn out to locate them in. In addition, observers are identified with entire time-extended histories, or diverging worlds, bundled together in the universal wavefunction. Of course, for a hypothetical bird's-eye observer, there is no uncertainty regarding the unfolding of the wavefunction - the \textbf{incoherence problem} does not exist for them. But for any branch-localized observer, there is justifiable self-locating uncertainty regarding which branch they are on. 

The ontological postulate is intimately connected to the statistical postulate, in the following sense:

\emph{Individual measurements supervene on the ontology. A self-locating data point is an ontic sample.}

Probability theories seek to assign rational degrees of belief to propositions about the world. They are designed to tell us the degrees of belief we \emph{ought} to have. The weakest sense in which one can derive an `ought' from an `is' is the following:

\emph{One ought to believe in what is the case.}

What then, can be said about degrees of belief? Can something be true about the world, to a degree? 

On the face of it, this doesn't seem plausible. However, in statements which refer to statistics of physical measurements, it can be true that the proposition $A$ is true in a proportion $x$ of those measurements. Since those measurements are direct instantiations of the underlying ontology, and \textbf{wavefunction monism} is true, there is a meaningful notion of a degree of belief about the truth of $A$. It is true in a proportion of wavefunction equal to $x$. 

I therefore put forward the following explication of  \textbf{center indifference} with respect to \textbf{units of wavefunction}, which has its roots in the classical approach to probability discussed in Section \ref{sec:equivocation_norms}:

\begin{tcolorbox}
\textbf{Definition 1 (Quantum probability)} 
\emph{In a temporal region of the wavefunction defined by some set of constraints, process $A$ has probability $\textrm{p}(A)=x$ if and only if $A$ occurs in a fraction $x$ of the total available wavefunction.}
\end{tcolorbox}

The only information an observer can glean which is relevant to their problem of self-location is the specification of boundary constraints which delineate their region of the wavefunction. To satisfy the implication in Eq. \eqref{eq:implication}, I now make the normative assertion that a rational observer should set their self-location uncertainty equal to the quantum probability. There are no further constraints on rational degrees of belief. 

\subsection{Event symmetry and representational models}

Our physical theories should contain more than 
the bare minimum required for calculations of data sets - they should, to the best of our knowledge, be representational of reality itself. In practice this means that, in addition to logical and ontological parsimony, we wish to encode true facts about reality into our representational models (\cite{ridley2025time}). Moreover, we should accept at face value what the model is telling us if it reduces the number of independent mathematical propositions (or more pertinently, \emph{types} of proposition) in our core physical theory, whilst retaining predictive power.

In Section \ref{sec:Time} I noted that, in the standard approach to quantum mechanics, time is not a dynamical variable. Rather, it is treated as a background parameter of the theory. This way of viewing time stands in stark contrast with the picture of relativistic spacetime theories, in which time is treated as one coordinate in a unified spacetime geometry. The geometric concept of time is \textbf{eternalist} - it includes all moments, or spacetime events, in all relativistic reference frames (\cite{horwitz_two_1988}). 

Recent work suggests that, rather than building a general-relativistic theory of quantum mechanics in the usual language of quantum systems, efforts should be refocused on a quantum theory which retains the event-based ontology of general relativity (\cite{maccone_fundamental_2019,giovannetti2023geometric}). For that to be possible, one must formalize the notion of an event within quantum mechanics. And one should do so in a way which retains the core properties of an event in spacetime theories. I now enumerate two of those properties:

\begin{tcolorbox}
\textbf{Definition 2 (Future-Past Symmetry)} - \emph{from a single event $\left\langle x,t\right\rangle$, the future and past lightcones extend symmetrically. Regardless of what is taken to be the `true' direction of time, there is an identical constraint on those events in the future and past of $\left\langle x,t\right\rangle$, which is just the point where all world lines $\left\langle \left\langle x_{1},t_{1}\right\rangle ,\ldots,\left\langle x,t\right\rangle ,\ldots,\left\langle x_{N},t_{N}\right\rangle \right\rangle$ connecting $N$ spacetime points to $\left\langle x,t\right\rangle$ \emph{cross}.}
\end{tcolorbox}

The presence of the event $\left\langle x,t\right\rangle$ in a sequence is a boundary condition which \emph{constrains} the other events that can be observed in sequence. From the bird's-eye, atemporal view of the entire spacetime, it can be viewed as a \emph{source/sink} for forwards-time directed processes in the future/past lightcones, and a \emph{sink/source} for backwards-time directed processes in the future/past lightcones:
\emph{Events correspond to crossing points of world lines which constrain the future and past.}
In conventional special relativity there is retrocausality in the sense that an event \emph{constrains what is possible} in the future and past of that event, and does so symmetrically.

\begin{tcolorbox}
\textbf{Definition 3 (Event Symmetry)} - \emph{the fundamental description of nature is independent of event location.}
\end{tcolorbox}

\textbf{Event symmetry} is a desirable property of any representational model of reality. Just as \textbf{strong neutral realism} requires there be no privileged observers, so there should be no privileged spacetime points. This principle is developed and argued for at length in \cite{ridley2025time}. 
 
\subsection{Fixed points on the Keldysh contour}

In the Keldysh formulation of dynamics, each instant of time corresponds to two time orientations, and it is not possible to neglect the forwards or backwards part of the time propagation in the `future' or `past' of some measurement time $t$ without losing essential physical information. In this way there is no way to distinguish between the two directions of propagation, or between different instants of time, such that both \textbf{Future-Past Symmetry} and \textbf{Event Symmetry} are automatically respected. 

\begin{figure}
\centering
  \includegraphics[width=.9\linewidth]{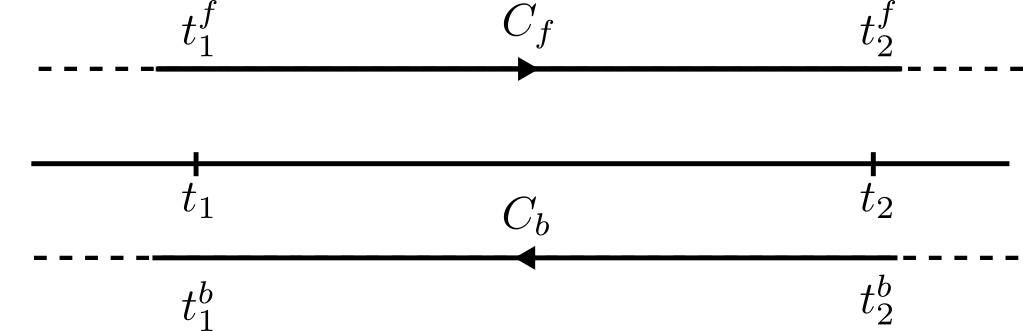}
  \caption{The Keldysh time contour in the time interval $\left[t_{1},t_{2}\right]$.}
  \label{fig:Keldysh_Contour}
\end{figure}

In the 1960s, it was independently noticed by Schwinger and Keldysh that quantum statistical expectation values are products of pairs of amplitudes describing processes with opposite time orientations (\cite{schwinger1961brownian,keldysh_diagram_1964}). This lead them to formulate many-body perturbation theory (MBPT) on a generalized time-contour composed of two copies of the time domain $[t_{1},t_{2}]$, as shown in Fig. \ref{fig:Keldysh_Contour}.
In this figure, forwards-time propagation is given in terms of times $t^{f}$ on the `upper' branch $C_{f}$ running from $t_{1}$ to $t_{2}$, and backwards-time propagation is described on the `lower' branch $C_{b}$ running from $t_{2}$ to $t_{1}$, with corresponding time labels $t_{1}^{f/b}$ and $t_{2}^{f/b}$ for the forwards/backwards-orientated time branch. The full Keldysh time contour, shown in Fig. \ref{fig:Keldysh_Contour}, is the direct sum of the two branches $C=C_{f} \oplus C_{b}$ (\cite{stefanucci2025nonequilibrium}). 

At this point, one can  implement the ETNU philosophy discussed in Section \ref{sec:Time} \emph{at every point on the Keldysh contour}, i.e. the state space of the system is replicated at each time \emph{and} at each time orientation, extending the state spaces found in the consistent histories framework or the multiple-time state formalism. This is done by assigning a distinct Hilbert space to each contour time, $\mathcal{H}_{t_{i}}^{\alpha}$, where $\alpha\in\left\{ f,b\right\}$ denotes the upper or lower branch of $C$. Then, denoting contour position by the variable $z$, one can define a Fock-like space of events, which is one of the primary desiderata stipulated by \cite{maccone_fundamental_2019} for a quantum theory of events:

\begin{tcolorbox}
\textbf{Definition 4 (Contour Space)} 

\begin{gather}\label{eq:event_space}
    \mathcal{H}_{C}=\mathbb{C}\oplus\overset{\infty}{\underset{N_{t}=1}{\bigoplus}}\overset{N_{t}}{\underset{i=1}{\bigotimes}}\int_{C}^{\oplus}\mathcal{H}_{z_{i}}dz_{i}\\
    =\mathbb{C}\oplus \int_{C}^{\oplus}\mathcal{H}_{z_{1}}dz_{1}\oplus\int_{C}^{\oplus}\mathcal{H}_{z_{1}}\otimes\mathcal{H}_{z_{2}}dz_{1}dz_{2}\oplus\ldots
\end{gather}
\end{tcolorbox}

In Eq. \eqref{eq:event_space}, the symbol $\int_{C}^{\oplus}$ denotes a direct integral over all times on $C$ (\cite{wils1970direct}). The universal wavefunction is defined on $\mathcal{H}_{C}$ in a representation-independent way, as a summation over 0-time, 1-time, 2-time...structures:

\begin{tcolorbox}
\textbf{Ontological postulate} 

\begin{gather}\label{eq:Psi_Universe}
    \left|\Psi_{U}\right\rangle=\left|0\right\rangle+\overset{\infty}{\underset{N_{t}=1}{\Sigma}}\overset{N_{t}}{\underset{i=1}{\otimes}}\int_{C}\left|\psi_{i}\right\rangle dz_{i}
\end{gather}
\end{tcolorbox}

Given an ordering of $N_{t}$ times $t_{N_{t}}>t_{N_{t}-1}>...>t_{1}$, there are two corresponding causal orderings, one on each branch of $C \equiv C_{b} \oplus C_{f}$:
\begin{gather}\label{eqn:f_b_ordering}
t_{N_{t}}^{f}>_{C}t_{N_{t}-1}^{f}>_{C}\ldots>_{C}t_{1}^{f}\\
t_{N_{t}}^{b}<_{C}t_{N_{t}-1}^{b}<_{C}\ldots<_{C}t_{1}^{b}
\end{gather}

where the contour-ordering notation $>_{C}$, $<_{C}$ is introduced as in \cite{stefanucci2025nonequilibrium}. This is the main innovation of the Keldysh contour: ordering in time is distinct from causal ordering, since causal influences propagate in the antichronological direction on the lower branch $C_{b}$. 

Each of the $N_{t}$ times in a history possesses two associated Hilbert spaces for the $f$ and $b$ components. Hence, the universal wavefunction has $2N_{t}$ temporal degrees of freedom and is a member of the following subspace of $\mathcal{H}_{C}$:
\begin{equation}
  \mathcal{H}_{C}\left(N_{t}\right)=\mathcal{H}_{t_{N_{t}}}^{b}\otimes\mathcal{H}_{t_{N_{t}}}^{f}\otimes\ldots\otimes\mathcal{H}_{t_{1}}^{b}\otimes\mathcal{H}_{t_{1}}^{f}
  \label{eqn:Hilbert_Space}
\end{equation}  

A wavefunction in this space is not defined at a single fixed `present', but at a sequence of moments with oppositely oriented  temporal parts acting as `source' or `sink' states for processes on the branches $C_{f}$ and $C_{b}$. 

I now introduce the second core postulate, encoding the transfer of physical information between states defined on $C$:

\begin{tcolorbox}
\textbf{Dynamical postulate}

\textit{The time derivative of the wavefunction at each point on $C$ is given by the TDSE:}
\begin{equation}\label{eq:Branch_TDSE}
i\hbar\partial_{t^{\alpha}}\left|\Psi^{\alpha}\left(t^{\alpha}\right)\right\rangle =\hat{H}^{\alpha}\left(t^{\alpha}\right)\left|\Psi^{\alpha}\left(t^{\alpha}\right)\right\rangle 
\end{equation}
\end{tcolorbox}

For all closed quantum systems the Hamiltonian operator is branch-independent, i.e. it takes on values on the upper/lower branches which are equal for the same physical time, $\hat{H}^{b}\left(t^{b}\right)=\hat{H}^{f}\left(t^{f}\right)$, and thus the dynamics respects \textbf{time symmetry}. For simplicity, indices on time arguments can therefore be dropped, $\left|\Psi^{\alpha}\left(t^{\alpha}\right)\right\rangle\equiv\left|\Psi^{\alpha}\left(t\right)\right\rangle$. 

The TDSE in Eq. \eqref{eq:Branch_TDSE} leads to a unitary mapping $U^{\alpha}\left(t_{2},t_{1}\right):\mathcal{H}_{t_{1}}^{\alpha}\mapsto\mathcal{H}_{t_{2}}^{\alpha}$ between the Hilbert spaces of different times on a single branch $\left|\Psi^{\alpha}\left(t_{2}\right)\right\rangle =U^{\alpha}\left(t_{2},t_{1}\right)\left|\Psi^{\alpha}\left(t_{1}\right)\right\rangle$, where $U^{\alpha}\left(t_{2},t_{1}\right)\equiv U^{\alpha}\left(t_{2}^{\alpha},t_{1}^{\alpha}\right)$ has the form (\cite{stefanucci2025nonequilibrium})
\begin{equation}
U^{\alpha}\left(t_{2},t_{1}\right)=\hat{\textrm{T}}_{C}\exp\left[-\frac{i}{\hbar}\int_{t_{1}^{\alpha}}^{t_{2}^{\alpha}}d\tau \hat{H}^{\alpha}\left(\tau\right)\right]
\label{eq:Branch_Propagator}
\end{equation}

and $\hat{\textrm{T}}_{C}$ orders operators chronologically (latest to the left) on $C_{f}$, and anti-chronologically on $C_{b}$. Note the formal similarity with Eq. \eqref{eq:propagator} in the consistent histories approach.

We now formalize the notion of a \emph{fixed point} in Contour Space, which corresponds to the intuitive notion of an event as a point in time at which distinct quantum histories coincide. 

\begin{tcolorbox}

\textbf{Definition 6 (Fixed Point) }

\textit{A fixed point at time $t$ is a temporal part of the wavefunction in the $\mathcal{H}_{t}^{b}\otimes\mathcal{H}_{t}^{f}$ subspace, with equal $f$ and $b$ parts.}
\end{tcolorbox}

Note that this definition allows for the retention of \textbf{Future-Past Symmetry}. Given a preparation of the state $\left|\psi\right\rangle$ of a system at some time $t_{1}$, all quantum histories in $\left|\Psi_{U}\right\rangle$ consistent with this preparation are constrained regardless of the contour branch. So there is a fixed point state at $t_{1}$, which is denoted:
\begin{equation}
\left\llbracket \psi\right\rrbracket_{t_{1}} \equiv\left|\psi^{b}\left(t_{1}\right)\right\rangle \otimes\left|\psi^{f}\left(t_{1}\right)\right\rangle
\label{eqn:1FP_Compact}
\end{equation}

\begin{figure}
\centering
    \includegraphics[width=.85\linewidth]{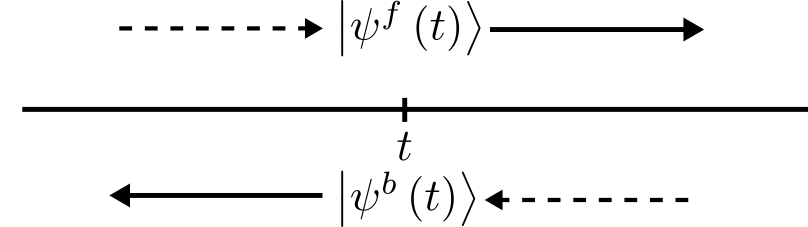}
  \caption{A single fixed point on the Keldysh contour.}
  \label{fig:1FP}
\end{figure}

This corresponds to an event in which the state is specified at $t_{1}$ (or a time-indexed projection, in the consistent histories language). The fixed point state connects to and constrains other points on $C$ in both time directions, in accordance with Eq. \eqref{eq:Branch_TDSE}. It is represented on $C$ in Fig. \ref{fig:1FP}: the forward-directed part of the fixed point defined at $t$ constrains times occurring `later' than $t^{f}$ on $C_{f}$, and the backward-directed part constrains times occurring `later' than $t^{b}$ on $C_{b}$. Each fixed point is connected to four temporal regions: it acts as a `source' of wavefunction in both time directions (the thick black arrows on Fig. \ref{fig:1FP}), and a `sink' for parts of the wavefunction constrained by times lying `earlier' on $C$ (dashed lines on Fig. \ref{fig:1FP}). Thus, for a full description of a measurement connecting times across the region $\left[t_{1},t_{2}\right]$ at least two fixed points are required, $N_{t}\geq2$ in Eq. \eqref{eq:Psi_Universe}. A quantum history sequence is defined in these terms:

\begin{tcolorbox}
\textbf{Definition 7 (Quantum history)}

\textit{A quantum history $h_{\mathbf{k}}$ extending across the time range $\left[t_{1},t_{2}\right]$ is a product state constructed from a sequence ${\mathbf{k}}=\left\langle k_{1},...,k_{N_{t}}\right\rangle $ of $N_{t}\geq2$ fixed points
\begin{equation}\label{eq:Quantum_History}
    h_{\mathbf{k}}=\overset{N_{t}}{\underset{i=1}{\otimes}}\left\llbracket \psi_{k_{i}}\right\rrbracket _{t_{i}}
\end{equation}
connected by unitary mappings and bounded by fixed points at $t_{1}$ and $t_{2}$.}
\end{tcolorbox}

In Eq. (\ref{eq:Quantum_History}), each $k_{i}$ in a history $h_{\mathbf{k}}$ ranges over a complete basis set spanning $\mathcal{H}_{t_{i}}^{\alpha}$. To allow us to apply the usual rules of probabilistic reasoning to quantum histories, we define a \emph{family} of quantum histories $\mathcal{F}_{H}$ by imposing the consistency condition that any pair of histories in a family $\left\{ \left|h_{\mathbf{k}}\right\rangle \right\}$ must be mutually orthogonal in the Keldysh space (although they may coincide at specific times):
\begin{equation}\label{eq:Quantum_History_Consistency}
    \left\langle h_{\mathbf{l}}\right.\left|h_{\mathbf{k}}\right\rangle =\delta_{\mathbf{kl}},
\end{equation}

where $\mathbf{k}\neq\mathbf{l}$ if $\left\llbracket \psi_{k_{i}}\right\rrbracket _{t_{i}}\neq\left\llbracket \psi_{l_{i}}\right\rrbracket _{t_{i}}$ for at least one value of $i \in \left[1,...,N_{t}\right]$. Note that the consistency condition Eq. (\ref{eq:Quantum_History_Consistency}) prevents the overlap of histories composed of different numbers of times $N_{t}$.

It is important to note that the universal wavefunction is not a superposition of histories, because the histories in a given family do not form a complete basis. Rather, they define the region of the universal wavefunction which is accessible to the observer when the observer is constrained by a sequence of $N_{t}$ projections or strong measurements, and therefore which is relevant for assignments of self-locating probabilities.

I have not defined families of histories with respect to a decoherence functional. This is intentional, but may come under criticism from those who appeal to decoherence theory in order to define a preferred quasi-classical basis for time-extended history states (\cite{gell_classical_1993,zurek1993preferred,gell_quantum_1996,zurek2003decoherence}). In my view, the decoherence program can only take us so far in describing the emergence of classical phenomena. No less relevant to this question are (i) the huge advances being currently made in condensed matter physics, materials science and theoretical chemistry, where highly detailed predictive multiscale models connecting the quantum many-body problem to macroscopic observables are under constant investigation (\cite{stefanucci2025nonequilibrium}), and (ii) the development of increasingly faster algorithms implemented on both classical (\cite{schleder2019dft}) and quantum (\cite{bauer2020quantum}) computing architectures for simulating real materials based on the models in (i). This constructive, multi-disciplinary effort offers at least a partial solution to the emergence and characterization of stable classical worlds, held together by spatially local interactions.

The fixed point model of an event treats events as constraints obeying \textbf{Future-Past Symmetry}, in that they act as both a source and a sink for both the future and past. In this sense, histories composed of fixed points exhibit retrocausality, but without the clean separation into future and past-oriented influences that is seen, for instance, in the TSVF. This model exhibits `all-at-once' retrocausality, in the sense that there is mutual causation between parts of the wavefunction located on the upper and lower branches of the Keldysh contour. The forwards and backwards oriented temporal parts of the wavefunction are inseparable components of the ontology.

\subsection{The Born rule replaced with the Vaidman rule}

Following the terminology of \cite{vaidman_schizophrenic_1998}, the \emph{measure of existence} of a history is now defined as the relative size of the wavefunction region occupied by that history.

\begin{tcolorbox}

\textbf{Definition 8 (Measure of existence)}

\textit{The measure of existence $m\left(h_{\textbf{k}}\right)$ of a quantum history $h_{\mathbf{k}}$ containing $N_{t}$ fixed points in the time range $\left[t_{1},t_{2}\right]$, is the ratio of the integral of the wavefunction $\triangle\Psi_{\textbf{k}}$ along this history, to that of all histories 
\begin{equation}\label{eq:MOE}
    m\left(h_{\textbf{k}}\right)=\frac{\ensuremath{\triangle\Psi_{\textbf{k}}}}{\underset{\mathbf{k'}}{\sum}\ensuremath{\triangle\Psi_{\textbf{k'}}}}
\end{equation}
consistent with the fixed point constraints.
}
\end{tcolorbox}

Fixed point constraints are imposed by taking the inner product of the integrated wavefunction with the `sink' state defined at the upper limits of the $2(N_{t}-1)$ segment integrals. Definition 8 gives precise meaning to the fraction of wavefunction connecting distinct events, and therefore (by Definition 1) a foundation for quantum probability:

\begin{tcolorbox}
\textbf{Statistical postulate (Vaidman rule)}
\textit{The quantum probability of a quantum history is equal to its measure of existence in the universal wavefunction.} 
\end{tcolorbox}

No explicit formula has yet been assumed for the measure of existence. The Vaidman rule is a \emph{conceptual} postulate, which is weaker than postulating the mathematical form of the Born measure outright, but which is sufficiently \emph{operational} to allow for a derivation of the Born rule in what follows. It also encodes a very tight connection between ontology and probability which gives \emph{physical meaning} to the latter, in accordance with Hartle's criterion for a derivation (\cite{hartle2021we}). 

I emphasize that the notion of quantum probability presented in this statistical postulate is a precisification of the idea, developed at length in Sections \ref{sec:equivocation_norms} and \ref{sec:self-locating}, that rational degrees of belief should respect \textbf{center indifference} with respect to \textbf{units of wavefunction}. One might object that, on face value, the Vaidman rule is not framed as a normative constraint on rational degrees of belief, but as a descriptive connection between `quantum probability' and regions of reality. However, the rational imperative is present in the notion of `probability' developed at length above, and in the discussion which opens Section \ref{sec:FPF}: if \textbf{wavefunction monism} is true, then our degree of belief in $x$ should equal the proportion of accessible reality in which $x$ is true. I will now derive the mathematical form of quantum probability from the Vaidman rule.

Consider a generic $N_{t}$ time sequence, $t_{1}<t_{2}<...<t_{N_{t}}$, at which there are $N_{t}$ fixed points. Some subset of these at times ${ t_{i_{1}}, t_{i_{2}},...,t_{i_{S_{t}}} }$ of cardinality $S_{t}<N_{t}$ corresponding to fixed points at which the wavefunction is known; these act as constraints on the quantum histories considered in the evaluation of Eq. \eqref{eq:MOE}. For ease of notation, assume the multiple-time state to be given by a member of some complete basis

\begin{equation}
    \Psi = \llbracket \phi_{i_{N_t}}\rrbracket_{t_{N_t}} \otimes...\otimes \llbracket \phi_{i_{1}}\rrbracket_{t_{1}}\equiv\mathbf{h}\left(N_{t}\right)
\end{equation}

where $\Psi \equiv \Psi(t^{b}_{N_{t}},t^{f}_{N_{t}},...,t^{b}_{1},t^{f}_{1})$ is a multiple-time function defined on $C$. Next, vary $\Psi$ across the temporal interval $[t_{1},t_{N_{t}}]$ along $C$:

\begin{equation}
    d\Psi=\frac{\partial\Psi}{\partial t_{1}^{f}}dt_{1}^{f}+\underset{i=2}{\overset{N_{t}-1}{\sum}}\left(\frac{\partial\Psi}{\partial t_{i}^{f}}dt_{i}^{f}+ \frac{\partial\Psi}{\partial t_{i}^{b}}dt_{i}^{b}\right)+\frac{\partial\Psi}{\partial t_{N_{t}}^{b}}dt_{N_{t}}^{b}
\end{equation}

The total change in wavefunction is then computed by taking the line integral along the following path on $C$

\begin{equation}\label{eq:integral_path}
\left(t_{N_{t}}^{b},t_{N_{t}}^{f},...,t_{1}^{b},t_{1}^{f}\right)\rightarrow\left(t_{N_{t}}^{b},t_{N_{t}}^{f},...,t_{1}^{b},t_{2}^{f}\right)\rightarrow...\rightarrow\left(t_{N_{t}-1}^{b},t_{N_{t}}^{f},...,t_{1}^{b},t_{2}^{f}\right),	
\end{equation}

shifting each time by one step in the forwards direction on $C^f$, and then by one step backwards on the lower branch $C^b$. The total change in wavefunction $D\Psi$ along this path is given explicitly by 

\begin{gather}
    D\Psi	=\overset{t_{2}^{f}}{\underset{t_{1}^{f}}{\int}}\frac{\partial\Psi}{\partial x}\left(t_{N_{t}}^{b},...,x\right)dx+...+\overset{t_{i+1}^{f}}{\underset{t_{i}^{f}}{\int}}\frac{\partial\Psi}{\partial x}\left(t_{N_{t}}^{b},...,t_{i}^{b},x,...,t_{2}^{f}\right)dx \nonumber \\
	+...+\overset{t_{N_{t}}^{f}}{\underset{t_{N_{t}-1}^{f}}{\int}}\frac{\partial\Psi}{\partial x}\left(t_{N_{t}}^{b},x,...,t_{2}^{f}\right)dx+\overset{t_{N_{t}-1}^{b}}{\underset{t_{N_{t}}^{b}}{\int}}\frac{\partial\Psi}{\partial x}\left(x,...,t_{2}^{f}\right)dx \nonumber \\
	+...+\overset{t_{i-1}^{b}}{\underset{t_{i}^{b}}{\int}}\frac{\partial\Psi}{\partial x}\left(t_{N_{t}-1}^{b},...,x,t_{i+1}^{f},...,t_{2}^{f}\right)dx+...+\overset{t_{1}^{b}}{\underset{t_{2}^{b}}{\int}}\frac{\partial\Psi}{\partial x}\left(t_{N_{t}-1}^{b},...,x,t_{2}^{f}\right)dx \label{eq:multitime_integral},
\end{gather}

where the integral bounds and arguments of the integrands follow the sequence in Eq. \eqref{eq:integral_path}. Each term in Eq. \eqref{eq:multitime_integral} is then evaluated explicitly as follows:

\begin{gather} D\Psi=U\left(t_{2}^{f},t_{1}^{f}\right)\Psi-\Psi+...+U\left(t_{i+1}^{f},t_{i}^{f}\right)...U\left(t_{2}^{f},t_{1}^{f}\right)\Psi-U\left(t_{i}^{f},t_{i-1}^{f}\right)...U\left(t_{2}^{f},t_{1}^{f}\right)\Psi \nonumber \\
+...+U\left(t_{N_{t}}^{f},t_{N_{t}-1}^{f}\right)...U\left(t_{2}^{f},t_{1}^{f}\right)\Psi-U\left(t_{N_{t}-1}^{f},t_{N_{t}-2}^{f}\right)...U\left(t_{2}^{f},t_{1}^{f}\right)\Psi \nonumber\\
+U\left(t_{N_{t}-1}^{b},t_{N_{t}}^{b}\right)U\left(t_{N_{t}}^{f},t_{N_{t}-1}^{f}\right)...U\left(t_{2}^{f},t_{1}^{f}\right)\Psi-U\left(t_{N_{t}}^{f},t_{N_{t}-1}^{f}\right)...U\left(t_{2}^{f},t_{1}^{f}\right)\Psi \nonumber\\
+...+U\left(t_{i-1}^{b},t_{i}^{b}\right)...U\left(t_{N_{t}-1}^{b},t_{N_{t}}^{b}\right)U\left(t_{N_{t}}^{f},t_{N_{t}-1}^{f}\right)...U\left(t_{2}^{f},t_{1}^{f}\right)\Psi \nonumber\\
    -U\left(t_{i}^{b},t_{i+1}^{b}\right)...U\left(t_{N_{t}-1}^{b},t_{N_{t}}^{b}\right)U\left(t_{N_{t}}^{f},t_{N_{t}-1}^{f}\right)...U\left(t_{2}^{f},t_{1}^{f}\right)\Psi \nonumber\\ 
+...+U\left(t_{1}^{b},t_{2}^{b}\right)...U\left(t_{N_{t}-1}^{b},t_{N_{t}}^{b}\right)U\left(t_{N_{t}}^{f},t_{N_{t}-1}^{f}\right)...U\left(t_{2}^{f},t_{1}^{f}\right)\Psi \nonumber\\    -U\left(t_{2}^{b},t_{3}^{b}\right)...U\left(t_{N_{t}-1}^{b},t_{N_{t}}^{b}\right)U\left(t_{N_{t}}^{f},t_{N_{t}-1}^{f}\right)...U\left(t_{2}^{f},t_{1}^{f}\right)\Psi.
\end{gather}

After cancellations, this expression simplifies to

\begin{equation}
    D\Psi=U\left(t_{1}^{b},t_{2}^{b}\right)...U\left(t_{N_{t}-1}^{b},t_{N_{t}}^{b}\right)U\left(t_{N_{t}}^{f},t_{N_{t}-1}^{f}\right)...U\left(t_{2}^{f},t_{1}^{f}\right)\Psi-\Psi.
\end{equation}

We then impose the boundary conditions at the `upper' end of the line integral by taking the inner product of $D\Psi$ with 

\begin{equation}
    \left|\phi_{i_{N_{t}-1}}^{b}\left(t_{N_{t}-1}\right)\right\rangle \left|\phi_{i_{N_{t}}}^{f}\left(t_{N_{t}}\right)\right\rangle ...\left|\phi_{i_{k-1}}^{b}\left(t_{k-1}\right)\right\rangle \left|\phi_{i_{k+1}}^{f}\left(t_{k+1}\right)\right\rangle ...\left|\phi_{i_{1}}^{b}\left(t_{1}\right)\right\rangle \left|\phi_{i_{2}}^{f}\left(t_{2}\right)\right\rangle,
\end{equation}

and using the fact that $\left\langle \phi_{i_{k}}^{\alpha}\left(t_{k}\right)\right|\left.\phi_{i_{k}}^{\alpha}\left(t_{l}\right)\right\rangle =\delta_{kl}$. Dropping branch indices from the resulting scalar product gives 

\begin{equation}
    \triangle\Psi=\left|\left\langle \phi_{i_{N_{t}-1}}\left(t_{N_{t}-1}\right)\right|U\left(t_{N_{t}-1},t_{N_{t}}\right)\left|\phi_{i_{N_{t}}}\left(t_{N_{t}}\right)\right\rangle ...\left\langle \phi_{i_{1}}\left(t_{1}\right)\right|U\left(t_{1},t_{2}\right)\left|\phi_{i_{2}}\left(t_{2}\right)\right\rangle \right|^{2},
\end{equation}

which is then substituted into Eq. \eqref{eq:MOE} to obtain the measure of existence for the $N_t$-time quantum history:

\begin{equation}\label{eq:Nt_Born}
    m\left(\mathbf{h}\left(N_{t}\right)\right)=\frac{\triangle\Psi}{\underset{i_{k_{1}}...i_{k_{N_{t}-S_{t}}}}{\sum}\triangle\Psi\left(...i_{k_{1}}...i_{k_{N_{t}-S_{t}}}...\right)}
\end{equation}

In the denominator, we sum over all histories consistent with the $S_t$ fixed point constraints. To see that Eq. \eqref{eq:Nt_Born} is equivalent to the Born rule, simply take the case of $N_{t}=2$, $S_{t}=1$, with the state known at time $t_1$. The normalization is $\underset{i_{2}}{\sum}\left|\left\langle \phi_{i_{1}}\left(t_{1}\right)\right|U\left(t_{1},t_{2}\right)\left|\phi_{i_{2}}\left(t_{2}\right)\right\rangle \right|^{2}=1$, so that the measure of existence is just 

\begin{equation}
    m\left(\mathbf{h}\left(2\right)\right)=\left|\left\langle \phi_{i_{1}}\left(t_{1}\right)\right|U\left(t_{1},t_{2}\right)\left|\phi_{i_{2}}\left(t_{2}\right)\right\rangle \right|^{2},
\end{equation}

which is the Born rule. Thus the factor of 2 in the exponent of the probability law is a direct result of the integration along two branches of the Keldysh contour. The same can be demonstrated for measurements involving pre- and post-selection, as was proven in \cite{ridley2023quantum}.

It is now worth taking a moment to compare the approach described here to other attempts at a derivation in the literature. 

First of all, this derivation does not rely on a partitioning of the universe into spatial subsystems and associated quantum `facts', so it is more general than the approach of  \cite{zurek_environment-assisted_2003,zurek_probabilities_2005}.

Moreover, the FPF obtains the mathematical form of the Born rule from local temporal constraints applied to the universal wavefunction in an event-symmetric way: it does not rely on specially chosen initial conditions, unlike approaches to probability in Bohmian mechanics which add the quantum equilibrium hypothesis to the core postulates (\cite{durr_quantum_1992}); neither does it require specially-selected final conditions of the universe to explain observed statistics (\cite{aharonov_two-time_2017}).

The FPF does not modify the Schr{\"o}dinger equation to impose collapse dynamically and then reconstruct the Born rule from a phenomenologically imposed rate-of-collapse parameter (\cite{bassi2023collapse}); rather the local dynamical maps are linear and unitary. This again makes the FPF a more general (and generalizable) approach, as such dynamical collapse is not present in general relativity. It furthermore contains no element of dynamical randomness, and in this respect is consistent with deterministic relativistic theories.

Adlam has recently mounted a critique of the concept of the measure of existence, in Chapter 6 of \cite{adlam2025saving}. There, Adlam quotes the Vaidman rule as an example of merely 
\emph{"calling the mod-squared amplitude a `measure of existence"} (\cite{adlam2025saving}).
But this is a straw-man argument when applied to the Vaidman rule. It might be that Adlam's critique holds for the \emph{Born-Vaidman rule}, which does, as discussed in Section \ref{sec:measure_of_existence} indeed postulate the mod-squared amplitude form of the measure of existence directly. The Vaidman rule does not do this, but instead gives a way to \emph{calculate} the mod-square amplitude probabilities given the temporal structure of the FPF.

\subsection{Many retrocausal worlds}
\begin{figure}[htp]
\centering
    \begin{subfigure}{0.65\textwidth}
        \includegraphics[clip, width=\linewidth]{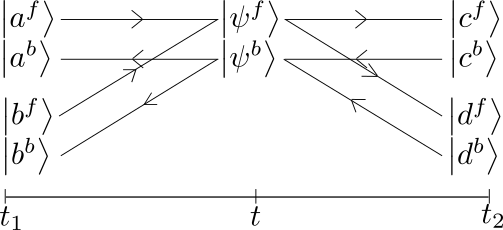}
        \subcaption{Schematic representation of the toy model considered in Eq. \eqref{eq:toy_universe}.} 
    \end{subfigure}%

    \begin{subfigure}{0.35\textwidth}
        \includegraphics[clip, width=\linewidth]{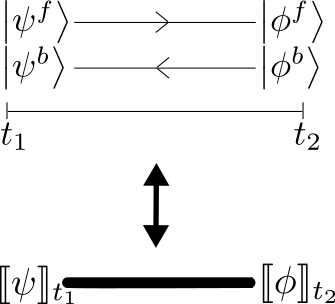}
        \subcaption{Compressed graphical representation of a two-time channel connecting the pair of fixed points $\left\llbracket \psi\right\rrbracket _{t_{1}}$ and $\left\llbracket \phi\right\rrbracket _{t_{2}}$.} 
    \end{subfigure}%
  \caption{}
  \label{fig:FPF_Sketch}
\end{figure}

We now come to a very rich picture of observable reality, which may be visualized as a network of fixed points constraining what can be observed in their future and their past. To fix ideas, I consider a toy model of a constrained region of the universe composed of a superposition of the following set of histories:
\begin{equation}\label{eq:toy_universe}
    \left\{ \left\llbracket c\right\rrbracket _{t_{2}},\left\llbracket d\right\rrbracket _{t_{2}}\right\} \otimes\left\llbracket \psi\right\rrbracket _{t}\otimes \left\{ \left\llbracket a\right\rrbracket _{t_{1}},\left\llbracket b\right\rrbracket _{t_{1}}\right\}
\end{equation}
which exist in the Hilbert space 

\begin{equation}
    \mathcal{H}_{t_{2}}^{b}\otimes\mathcal{H}_{t_{2}}^{f}\otimes\mathcal{H}_{t}^{b}\otimes\mathcal{H}_{t}^{f}\otimes\mathcal{H}_{t_{1}}^{b}\otimes\mathcal{H}_{t_{1}}^{f}
\end{equation}

where the fixed point states comprising the histories correspond to macroscopic states of the universe. This model is illustrated schematically in Fig. \ref{fig:FPF_Sketch} (a). Here, the fixed point $\left\llbracket \psi \right\rrbracket _{t}$ at the intermediate time $t \in \left[ t_{1},t_{2} \right]$ serves as a node for two branches in the future of $t$ and two branches in its past. Each connected pair of fixed points can be thought of as conjoined by a separate Keldysh contour, since each pairing is achieved via two branch channels. From here on, I  represent connected fixed points graphically using the compressed schematic in Fig. \ref{fig:FPF_Sketch} (b), and I refer to such objects as \emph{quantum history segments}.

We interpret this picture literally - in accordance with \textbf{extensive neutral realism}, all histories are simultaneously realized. What does this mean for the epistemic situation of an observer $O_{\psi}$ who observes the state to be $\left|\psi\right\rangle$ at time $t$? They seem to genuinely exist at the spacetime crossroads, with two pasts and two futures. From a bird's-eye view, they seem to have converged to the present moment from two distinct observers in the past, and apparently they are about to split into two new versions of themselves. This leads to new types of question: in addition to uncertainty about the future, how can they \emph{not} have uncertainty about their past? Should they not have a recollection of their distinct past selves? 

\begin{figure}[htp]
\centering
    \begin{subfigure}{\linewidth}
    \includegraphics[clip, width=1.0\linewidth]{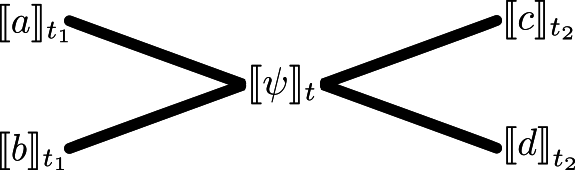}
    \subcaption{Overlapping worlds in the future and past of $t$.}
    \end{subfigure}
    \begin{subfigure}{\linewidth}
    \includegraphics[clip, width=1.0\linewidth]{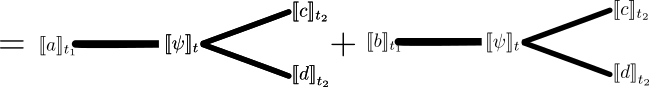}
    \subcaption{Overlapping worlds in the future of $t$, diverging worlds in the past of $t$.}
    \end{subfigure}
    \begin{subfigure}{\linewidth}
    \includegraphics[clip, width=1.0\linewidth]{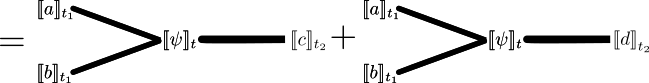}
    \subcaption{Diverging worlds in the future of $t$, overlapping worlds in the past of $t$.}
    \end{subfigure}
    \begin{subfigure}{\linewidth}
    \includegraphics[clip, width=1.0\linewidth]{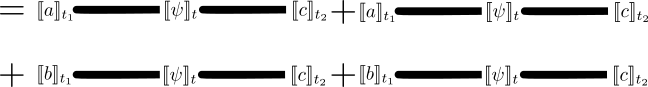}
    \subcaption{Diverging worlds in the future and past of $t$.}
    \end{subfigure}
    \caption{Formally equivalent decompositions of the toy universe specified in Eq. \eqref{eq:toy_universe}.}
    \label{fig:MRW_Semantics}
\end{figure}

At this point, I draw upon the work done in Section \ref{sec:overdiverge}. We must decide whether to identify the observer with their temporal parts, or with entire temporally-extended histories. The bird's-eye view is shown in Fig \ref{fig:MRW_Semantics} (a). A corresponding account of personal identity should be given, which makes sense of belief assignments about the future of $O_{\psi}$, and which can be reconciled with their experience of the past.

If I simply identify observer $O_{\psi}$ with its temporal part at $t$, then that person exists in distinct overlapping worlds both in the past and the future of $t$. I refer to this as the Many Overlapping Retrocausal Worlds (MORW) view.

The second possibility is illustrated in Fig. \ref{fig:MRW_Semantics} (b). Here, worlds are diverging in the past of $t$ but overlapping in its future. The notion of personal identity employed here is equivalent to two universes in the standard forwards-time MWI with overlapping worlds. I refer to this as the Many Many Worlds Future (MMWF) view.

We can also consider the case shown in Fig. \ref{fig:MRW_Semantics} (c), where worlds are overlapping in the past of $t$ but diverging in its future. This corresponds to two worlds in which the observer at times later than $t$ is a spacetime world-tube with memory of two different pasts. I refer to this as the Many Many Worlds Past (MMWP) view.

Finally, I come to the decomposition in Fig. \ref{fig:MRW_Semantics} (d), which represents the model as a bundle of four separate histories, in each of which the observer is identified with their entire spacetime world-tube. This I refer to as the Many Diverging Retrocausal Worlds (MDRW) view.

We can see that all the decompositions in Fig. \ref{fig:MRW_Semantics} are physically equivalent by considering the measure of existence for the entire toy model, which is the normalization factor for measures of individual histories. It can be decomposed as

\begin{gather}\label{MRW_MOE1}
    m\left(\left|\Psi_{U}\right\rangle \right)=\left(\left|\left\langle a\right|U\left(t_{1},t\right)\left|\psi\right\rangle \right|^{2}+\left|\left\langle b\right|U\left(t_{1},t\right)\left|\psi\right\rangle \right|^{2}\right)\left(\left|\left\langle \psi\right|U\left(t,t_{2}\right)\left|c\right\rangle \right|^{2}+\left|\left\langle \psi\right|U\left(t,t_{2}\right)\left|d\right\rangle \right|^{2}\right)
\end{gather}

i.e. all branches in the past glued to all branches in the future of $t$, corresponding to the MORW view shown in Fig. \ref{fig:MRW_Semantics} (a). We can also decompose it as a sum of single quantum history segments in the past of $t$ glued to both future segments, in the MMWF view shown in Fig. \ref{fig:MRW_Semantics} (b):

\begin{gather}
    m\left(\left|\Psi_{U}\right\rangle \right)=\left|\left\langle a\right|U\left(t_{1},t\right)\left|\psi\right\rangle \right|^{2}\left(\left|\left\langle \psi\right|U\left(t,t_{2}\right)\left|c\right\rangle \right|^{2}+\left|\left\langle \psi\right|U\left(t,t_{2}\right)\left|d\right\rangle\right|^{2}\right)\\
    +\left|\left\langle b\right|U\left(t_{1},t\right)\left|\psi\right\rangle \right|^{2}\left(\left|\left\langle \psi\right|U\left(t,t_{2}\right)\left|c\right\rangle \right|^{2}+\left|\left\langle \psi\right|U\left(t,t_{2}\right)\left|d\right\rangle\right|^{2}\right)\nonumber
\end{gather}

Likewise, one can decompose $m\left(\left|\Psi_{U}\right\rangle \right)$ as a sum of single quantum history segments in the future of $t$ glued to both past segments, in the MMWP view shown in Fig. \ref{fig:MRW_Semantics} (c):

\begin{gather}
    m\left(\left|\Psi_{U}\right\rangle \right)=\left(\left|\left\langle a\right|U\left(t_{1},t\right)\left|\psi\right\rangle \right|^{2}+\left|\left\langle b\right|U\left(t_{1},t\right)\left|\psi\right\rangle\right|^{2}\right)\left|\left\langle \psi\right|U\left(t,t_{2}\right)\left|c\right\rangle \right|^{2}\\
    +\left(\left|\left\langle a\right|U\left(t_{1},t\right)\left|\psi\right\rangle \right|^{2}+\left|\left\langle b\right|U\left(t_{1},t\right)\left|\psi\right\rangle \right|^{2}\right)\left|\left\langle \psi\right|U\left(t,t_{2}\right)\left|d\right\rangle \right|^{2}\nonumber
\end{gather}

Finally, the total measure of existence can be decomposed as the sum of existence measures of the four quantum histories taken separately, in the MDRW view shown in Fig. \ref{fig:MRW_Semantics} (d):

\begin{gather}\label{MRW_MOE_2}
    m\left(\left|\Psi_{U}\right\rangle \right)=\left|\left\langle a\right|U\left(t_{1},t\right)\left|\psi\right\rangle \left\langle \psi\right|U\left(t,t_{2}\right)\left|c\right\rangle \right|^{2}+\left|\left\langle a\right|U\left(t_{1},t\right)\left|\psi\right\rangle \left\langle \psi\right|U\left(t,t_{2}\right)\left|d\right\rangle \right|^{2}\\
    +\left|\left\langle b\right|U\left(t_{1},t\right)\left|\psi\right\rangle \left\langle \psi\right|U\left(t,t_{2}\right)\left|c\right\rangle \right|^{2}+\left|\left\langle b\right|U\left(t_{1},t\right)\left|\psi\right\rangle \left\langle \psi\right|U\left(t,t_{2}\right)\left|d\right\rangle \right|^{2}\nonumber
\end{gather}

We see, as a general rule, that \emph{quantum history segments that are simultaneous in time contribute additively to the measure of existence; consecutive quantum history segments are multiplied}.

On the MORW view, the experience of an observer described by this toy model is indeed unlike that of any person who has ever conducted an experiment, and they should have records of their multiple pasts. Since in practice they do not, I rule out this view. The MMWP view can be ruled out for the same reasons. The MMWF (and the MMWP) view divides the time contour into `past' and `future' regions, in which different concepts of personal identity obtain. Here, we can apply \textbf{Future-Past Symmetry} at the metaphysical level - there is no reason why our fundamental concept of what an observer is should be dependent on an arbitrary slicing of the universe into past, present and future. 

This leaves us with the MDRW viewpoint. The toy model is really just a bundle of histories which have an identical temporal part at $t$. Just as the incoherence problem has a solution with indexicalism, so too do the problems with backwards branching - there are \emph{four} copies of observer $O_\psi$ at time $t$, each localized to a distinct quantum history, or world-tube, with a distinct pair of quantum history segments in the future and past and distinct weighting assigned to the whole history. Each of these observers only has records of a single past observation at $t_{1}$, and they have genuine uncertainty regarding the outcome of the measurement they will perform at time $t_{2}$. There is an objective grounding for indexical uncertainty about which world-tube they are in, where world-tubes are conceived of as a sequence of physical states running in parallel through the time-extended quantum state, but not overlapping with other world-tubes. Retrocausality therefore seems to be easier to stomach with divergence/indexicalism than with branching/fission. 

As shown in Eqs. \eqref{MRW_MOE1}-\eqref{MRW_MOE_2}, the formalism does not distinguish between the MORW, MMWF, MMWP and MDRW pictures. However, the MDRW interpretation seems most plausible to us given its intuitive characterization of personal identity. In the real universe, the universal wavefunction contains huge numbers of distinct spacetime world-tubes, bundled together but taking different paths through its enormous network structure. These world-tubes may cross each other in the sense that they share fixed points. A history containing one fixed point need not be identified with all the other histories which share it. And their size relative to that of other quantum histories in the bundle is given by the Born measure. So as an individual observer who knows quantum mechanics, and in particular the FPF, they know they are identified with an entire history and they can compute the amount of reality which is allocated to each history. 

That this is the structure explaining the Born rule should now seem rather more persuasive - no further calculations or formal rules are required to compute the relative size of a quantum history. In accordance with the Vaidman rule, an observer simply needs to add up all the parts of the wavefunction which comprise their history and assume that their measurements are a statistical sample across the entire structure.

This is the deep lesson - the FPF, suitably interpreted with the MDRW view, allows us to do quantum mechanics with three bird's-eye postulates - the ontological, dynamical and composition postulates, and one local constraint on rational degrees of belief. So the Born measure does not need to be postulated separately as if it came from nowhere. It is very clear which physical property it corresponds to in the wavefunction. The wavefunction is positively drowning in Born measure. And we are rationally impelled to distribute degrees of belief over \textbf{units of wavefunction}.

I have argued in Sections \ref{sec:probability} and \ref{sec:probability_MWI} for a tightening of the connection between probability and ontology. Within the FPF, the mathematical structure of the Born rule does not have to be postulated or derived from auxiliary assumptions, rather it follows from the definition of probability, ontology and two-time dynamics.

\section{Conclusions}

In this paper, I have surveyed the major problems with understanding probability in the MWI, both at the conceptual and structural level. 

The conceptual work has two components. Firstly, I have argued that there is an historical precedent for a notion of probability as the relative proportion of physical reality. This embeds probability within the ontology of the theory, and satisfies the criterion that degrees of belief should be determined by physical structure, and not the other way around. Secondly, I have identified reasons why the divergence theory of identity is preferable to the branching view within the MWI, and that it is possible to make sense of self-locating probability based on the diverging worlds view.

At the structural level, I have argued that existing attempts to formally derive the Born measure within the MWI are unsatisfactory. The alternative strategy in the literature, of simply postulating the Born rule outright, gives us no insight into the physical origin of quantum probability. For this reason, and drawing upon independent arguments for the introduction of \textbf{event symmetry} into the quantum theory, I have argued for a modification of the ontic and dynamical content of quantum mechanics which tightens the connection between this structure and the physical meaning of the Born rule, thereby satisfying Hartle's criterion for a derivation (\cite{hartle2021we}). This strategy also lends \emph{operational} meaning to the concept of a \textbf{measure of existence}, which is \emph{calculated} to yield the Born measure and its generalizations. Furthermore, I have interpreted this formalism within an Everettian framework in which backwards branching is permitted, arguing that the concept of probability can be understood within this framework if we adopt the MDRW interpretation. 

I am optimistic that the ideas contained in this work can be connected to many open research questions, some of which I sketch here:

\begin{enumerate}
    \item Can the concept of quantum probability as wavefunction fractions be shown to reproduce statistical mechanical probabilities under the appropriate conditions? This seems highly plausible given recent work on thermalization and generalized Gibbs ensembles in many body systems (\cite{langen2015experimental}).
    \item Can an effective thermodynamic arrow of time be shown to emerge from the `mutual causation' picture of the FPF?
    \item The time-local fixed-point constraints serve as crossing points for large numbers of simultaneous histories. Can this fact be related to energy-time uncertainty? 
    \item Can the multiple-time, retrocausal model developed here satisfy new criteria for lawhood (\cite{adlam2022laws,adlam2022two}) in all-at-once retrocausal theories?
    \item Can the unitary, retrocausal resolution of the measurement problem be extended to relativistic systems? 
\end{enumerate}

Bearing all these questions in mind, and especially the last one, I would finally like to comment on the \textbf{generalizability} of the approach described here - the FPF renders quantum mechanics conceptually compatible with general relativity, since it retains determinism, an all-at-once block universe picture and a formal concept of an event. It also allows for the derivation of the Born rule. The price to pay for these striking advantages is the existence of many retrocausal worlds, bundled together in the universal wavefunction.

\bmhead{Acknowledgements}
I am grateful to Emily Adlam, Yakir Aharonov, Jonny Blamey, Eliahu Cohen, Philipp Strasberg, Riku Tuovinen and Lev Vaidman for many useful discussions. I also wish to thank Yuval Idan and Laurie Leterte for inspecting the manuscript during the revision process. Finally, I am grateful to the three anonymous reviewers for their insightful comments on this work. 

\section*{Declarations}

\bmhead{Funding}
This work has been supported by the European Union’s Horizon Europe research and innovation programme under grant agreement No. 101178170, by the Israel Science Foundation Grant No. 2064/19 and the National Science Foundation–US-Israel Binational Science Foundation Grant No. 735/18.


\begin{thebibliography}{}
\providecommand{\doi}[1]{\url{https://doi.org/#1}}
\bibcommenthead

\bibitem[\protect\citeauthoryear{Adlam}{Adlam}{2018}]{adlam2018spooky}
Adlam, E. 2018.
\newblock Spooky action at a temporal distance.
\newblock {\em Entropy\/}~{\em 20\/}(1): 41 .

\bibitem[\protect\citeauthoryear{Adlam}{Adlam}{2022a}]{adlam2022laws}
Adlam, E. 2022a.
\newblock Laws of nature as constraints.
\newblock {\em Foundations of Physics\/}~{\em 52\/}(1): 28 .

\bibitem[\protect\citeauthoryear{Adlam}{Adlam}{2022b}]{adlam2022two}
Adlam, E. 2022b.
\newblock Two roads to retrocausality.
\newblock {\em Synthese\/}~{\em 200\/}(5): 422 .

\bibitem[\protect\citeauthoryear{Adlam}{Adlam}{2025}]{adlam2025saving}
Adlam, E. 2025.
\newblock {\em Saving Science from Quantum Mechanics: The Epistemology of the Measurement Problem}.
\newblock Oxford University Press.

\bibitem[\protect\citeauthoryear{Aharonov, Bergmann, and Lebowitz}{Aharonov et~al.}{1964}]{aharonov_time_1964}
Aharonov, Y., P.G. Bergmann, and J.L. Lebowitz. 1964.
\newblock Time symmetry in the quantum process of measurement.
\newblock {\em Physical Review\/}~{\em 134\/}(6B): B1410 .

\bibitem[\protect\citeauthoryear{Aharonov, Cohen, and Landsberger}{Aharonov et~al.}{2017}]{aharonov_two-time_2017}
Aharonov, Y., E.~Cohen, and T.~Landsberger. 2017, March.
\newblock The {Two}-{Time} {Interpretation} and {Macroscopic} {Time}-{Reversibility}.
\newblock {\em Entropy\/}~{\em 19\/}(3): 111.
\newblock \doi{10.3390/e19030111} .

\bibitem[\protect\citeauthoryear{Aharonov, Popescu, and Tollaksen}{Aharonov et~al.}{2014}]{aharonov_each_2014}
Aharonov, Y., S.~Popescu, and J.~Tollaksen. 2014.
\newblock Each instant of time a new universe, {\em Quantum theory: a two-time success story},  21--36. Springer.

\bibitem[\protect\citeauthoryear{Aharonov, Popescu, Tollaksen, and Vaidman}{Aharonov et~al.}{2009}]{aharonov_multiple-time_2009}
Aharonov, Y., S.~Popescu, J.~Tollaksen, and L.~Vaidman. 2009.
\newblock Multiple-time states and multiple-time measurements in quantum mechanics.
\newblock {\em Physical Review A\/}~{\em 79\/}(5): 052110 .

\bibitem[\protect\citeauthoryear{Aharonov and Vaidman}{Aharonov and Vaidman}{1991}]{aharonov1991complete}
Aharonov, Y. and L.~Vaidman. 1991.
\newblock Complete description of a quantum system at a given time.
\newblock {\em Journal of Physics A: Mathematical and General\/}~{\em 24\/}(10): 2315 .

\bibitem[\protect\citeauthoryear{Aharonov and Vaidman}{Aharonov and Vaidman}{2008}]{aharonov_two-state_2008}
Aharonov, Y. and L.~Vaidman. 2008.
\newblock The two-state vector formalism: an updated review, {\em Time in quantum mechanics},  399--447. Springer.

\bibitem[\protect\citeauthoryear{Albert}{Albert}{1996}]{albert_elementary_1996}
Albert, D.Z. 1996.
\newblock Elementary quantum metaphysics, {\em Bohmian mechanics and quantum theory: {An} appraisal},  277--284. Springer.

\bibitem[\protect\citeauthoryear{Albert}{Albert}{2010}]{albert_probability_2010}
Albert, D.Z. 2010.
\newblock Probability in the {Everett} picture, {\em Many worlds},  355--368. Oxford University Press.

\bibitem[\protect\citeauthoryear{Albert}{Albert}{2013}]{albert2013wave}
Albert, D.Z. 2013.
\newblock Wave function realism, In {\em The wave function: Essays on the metaphysics of quantum mechanics},  eds. Albert, D.Z. and A.~Ney,  52--57. Oxford, UK: Oxford University Press.

\bibitem[\protect\citeauthoryear{Anderson}{Anderson}{2012}]{anderson2012problem}
Anderson, E. 2012.
\newblock Problem of time in quantum gravity.
\newblock {\em Annalen der Physik\/}~{\em 524\/}(12): 757--786 .

\bibitem[\protect\citeauthoryear{Araújo}{Araújo}{2019}]{araujo_probability_2019}
Araújo, M. 2019, March.
\newblock Probability in {Two} {Deterministic} {Universes}.
\newblock {\em Found Phys\/}~{\em 49\/}(3): 202--231 .

\bibitem[\protect\citeauthoryear{Barrett}{Barrett}{2017}]{barrett_typical_2017}
Barrett, J.A. 2017.
\newblock Typical worlds.
\newblock {\em Studies in History and Philosophy of Science Part B: Studies in History and Philosophy of Modern Physics\/}~58: 31--40 .

\bibitem[\protect\citeauthoryear{Bassi, Dorato, and Ulbricht}{Bassi et~al.}{2023}]{bassi2023collapse}
Bassi, A., M.~Dorato, and H.~Ulbricht. 2023.
\newblock Collapse models: a theoretical, experimental and philosophical review.
\newblock {\em Entropy\/}~{\em 25\/}(4): 645 .

\bibitem[\protect\citeauthoryear{Bassi, Lochan, Satin, Singh, and Ulbricht}{Bassi et~al.}{2013}]{bassi_models_2013}
Bassi, A., K.~Lochan, S.~Satin, T.P. Singh, and H.~Ulbricht. 2013.
\newblock Models of wave-function collapse, underlying theories, and experimental tests.
\newblock {\em Reviews of Modern Physics\/}~{\em 85\/}(2): 471 .

\bibitem[\protect\citeauthoryear{Bauer, Bravyi, Motta, and Chan}{Bauer et~al.}{2020}]{bauer2020quantum}
Bauer, B., S.~Bravyi, M.~Motta, and G.K.L. Chan. 2020.
\newblock Quantum algorithms for quantum chemistry and quantum materials science.
\newblock {\em Chemical reviews\/}~{\em 120\/}(22): 12685--12717 .

\bibitem[\protect\citeauthoryear{Bell}{Bell}{1990}]{bell_against_1990}
Bell, J. 1990.
\newblock Against ‘measurement’.
\newblock {\em Physics world\/}~{\em 3\/}(8): 33 .

\bibitem[\protect\citeauthoryear{Berenstain}{Berenstain}{2020}]{berenstain_privileged-perspective_2020}
Berenstain, N. 2020.
\newblock Privileged-perspective realism in the quantum multiverse, In {\em The Foundation of Reality: Fundamentality, Space, and Time},  eds. David~Glick, G.D. and A.~Marmodoro,  102–122. Oxford, UK: Oxford University Press.

\bibitem[\protect\citeauthoryear{Bertrand}{Bertrand}{1907}]{bertrand_calcul_1907}
Bertrand, J. 1907.
\newblock {\em Calcul des probabilités}.
\newblock Gauthier-Villars.

\bibitem[\protect\citeauthoryear{Blackshaw}{Blackshaw}{2023}]{blackshaw2023probability}
Blackshaw, N.I. 2023.
\newblock {\em Probability and Branching in Everettian Quantum Physics}.
\newblock Ph.\ D. thesis, University of Bristol.

\bibitem[\protect\citeauthoryear{Bohm}{Bohm}{1952}]{bohm_suggested_1952}
Bohm, D. 1952.
\newblock A suggested interpretation of the quantum theory in terms of" hidden" variables. {I}.
\newblock {\em Physical review\/}~{\em 85\/}(2): 166 .

\bibitem[\protect\citeauthoryear{Born}{Born}{1926}]{born1926quantenmechanik}
Born, M. 1926.
\newblock Quantenmechanik der sto{\ss}vorg{\"a}nge.
\newblock {\em Zeitschrift f{\"u}r physik\/}~{\em 38\/}(11-12): 803--827 .

\bibitem[\protect\citeauthoryear{Brown and Ben~Porath}{Brown and Ben~Porath}{2020}]{brown_everettian_2020}
Brown, H.R. and G.~Ben~Porath. 2020.
\newblock Everettian probabilities, the deutsch-wallace theorem and the principal principle, In {\em Quantum, Probability, Logic: Itamar Pitowsky's Work and Influence},  eds. Hemmo, M. and O.~Shenker,  165--198. Cham: Springer International Publishing.

\bibitem[\protect\citeauthoryear{Builes}{Builes}{2019}]{builes_self-locating_2019}
Builes, D. 2019.
\newblock Self-{Locating} {Evidence} and the {Metaphysics} of {Time}.
\newblock {\em Philosophy and Phenomenological Research\/}~{\em 99\/}(2): 478--490 .

\bibitem[\protect\citeauthoryear{Builes}{Builes}{2020}]{builes_time-slice_2020}
Builes, D. 2020.
\newblock Time-slice rationality and self-locating belief.
\newblock {\em Philosophical Studies\/}~{\em 177\/}(10): 3033--3049 .

\bibitem[\protect\citeauthoryear{Calosi}{Calosi}{2018}]{calosi_quantum_2018}
Calosi, C. 2018.
\newblock Quantum monism: an assessment.
\newblock {\em Philosophical Studies\/}~{\em 175\/}(12): 3217--3236 .

\bibitem[\protect\citeauthoryear{Carnap}{Carnap}{1945}]{carnap_two_1945}
Carnap, R. 1945.
\newblock The two concepts of probability: {The} problem of probability.
\newblock {\em Philosophy and phenomenological research\/}~{\em 5\/}(4): 513--532 .

\bibitem[\protect\citeauthoryear{Carroll and Sebens}{Carroll and Sebens}{2014}]{carroll_many_2014}
Carroll, S.M. and C.T. Sebens. 2014.
\newblock Many worlds, the born rule, and self-locating uncertainty, {\em Quantum theory: {A} two-time success story},  157--169. Springer.

\bibitem[\protect\citeauthoryear{Carroll and Singh}{Carroll and Singh}{2019}]{carroll_mad-dog_2019}
Carroll, S.M. and A.~Singh. 2019.
\newblock Mad-dog {Everettianism}: {Quantum} mechanics at its most minimal, {\em What is {Fundamental}?},  95--104. Springer.

\bibitem[\protect\citeauthoryear{Chen}{Chen}{2019}]{chen_realism_2019}
Chen, E.K. 2019.
\newblock Realism about the wave function.
\newblock {\em Philosophy compass\/}~{\em 14\/}(7): e12611 .

\bibitem[\protect\citeauthoryear{Cohen and Aharonov}{Cohen and Aharonov}{2017}]{cohen2017quantum}
Cohen, E. and Y.~Aharonov. 2017.
\newblock Quantum to classical transitions via weak measurements and post-selection, {\em Quantam Structural Studies: Classical Emergence from the Quantum Level},  401--425. World Scientific.

\bibitem[\protect\citeauthoryear{Colbeck and Renner}{Colbeck and Renner}{2012}]{colbeck_is_2012}
Colbeck, R. and R.~Renner. 2012.
\newblock Is a system's wave function in one-to-one correspondence with its elements of reality?
\newblock {\em Physical Review Letters\/}~{\em 108\/}(15): 150402 .

\bibitem[\protect\citeauthoryear{Cotler, Duan, Hou, Wilczek, Xu, Yin, and Zu}{Cotler et~al.}{2017}]{cotler2017experimental}
Cotler, J., L.M. Duan, P.Y. Hou, F.~Wilczek, D.~Xu, Z.Q. Yin, and C.~Zu. 2017.
\newblock Experimental test of entangled histories.
\newblock {\em Annals of Physics\/}~387: 334--347 .

\bibitem[\protect\citeauthoryear{Cunningham}{Cunningham}{2014}]{cunningham_branches_2014}
Cunningham, A.J. 2014.
\newblock Branches in the {Everett} interpretation.
\newblock {\em Studies in History and Philosophy of Science Part B: Studies in History and Philosophy of Modern Physics\/}~46: 247--262 .

\bibitem[\protect\citeauthoryear{Dawid and Th{\'e}bault}{Dawid and Th{\'e}bault}{2015}]{dawid2015many}
Dawid, R. and K.P. Th{\'e}bault. 2015.
\newblock Many worlds: decoherent or incoherent?
\newblock {\em Synthese\/}~{\em 192\/}(5): 1559--1580 .

\bibitem[\protect\citeauthoryear{de~Laplace}{de~Laplace}{1812}]{laplace_theorie_1812}
de~Laplace, P.S. 1812.
\newblock {\em Th{\'e}orie analytique des probabilit{\'e}s}.
\newblock Courcier.

\bibitem[\protect\citeauthoryear{Deffner}{Deffner}{2017}]{deffner2017demonstration}
Deffner, S. 2017.
\newblock Demonstration of entanglement assisted invariance on ibm's quantum experience.
\newblock {\em Heliyon\/}~{\em 3\/}(11): e00444 .

\bibitem[\protect\citeauthoryear{Deutsch}{Deutsch}{1985}]{deutsch_quantum_1985}
Deutsch, D. 1985.
\newblock Quantum theory as a universal physical theory.
\newblock {\em International Journal of Theoretical Physics\/}~{\em 24\/}(1): 1--41 .

\bibitem[\protect\citeauthoryear{Deutsch}{Deutsch}{1999}]{deutsch_quantum_1999}
Deutsch, D. 1999.
\newblock Quantum theory of probability and decisions.
\newblock {\em Proc. R. Soc. Lond. A\/}~{\em 455\/}(1988): 3129--3137 .

\bibitem[\protect\citeauthoryear{DeWitt}{DeWitt}{1967}]{dewitt1967quantum}
DeWitt, B.S. 1967.
\newblock Quantum theory of gravity. {I}. {T}he canonical theory.
\newblock {\em Phys. Rev.\/}~{\em 160\/}(5): 1113.
\newblock \doi{10.1103/PhysRev.160.1113} .

\bibitem[\protect\citeauthoryear{DeWitt}{DeWitt}{1970}]{dewitt_quantum_1970}
DeWitt, B.S. 1970.
\newblock Quantum mechanics and reality.
\newblock {\em Physics today\/}~{\em 23\/}(9): 30--35 .

\bibitem[\protect\citeauthoryear{Diosi}{Diosi}{1987}]{diosi1987universal}
Diosi, L. 1987.
\newblock A universal master equation for the gravitational violation of quantum mechanics.
\newblock {\em Physics letters A\/}~{\em 120\/}(8): 377--381 .

\bibitem[\protect\citeauthoryear{Drezet}{Drezet}{2021}]{drezet2021making}
Drezet, A. 2021.
\newblock Making sense of born’s rule $p_{\alpha}=\left\vert \psi_{\alpha}\right\vert^{2}$ with the many-minds interpretation.
\newblock {\em Quantum Studies: Mathematics and Foundations\/}~{\em 8\/}(3): 315--336 .

\bibitem[\protect\citeauthoryear{Dürr, Goldstein, Norsen, Struyve, and Zanghì}{Dürr et~al.}{2014}]{durr_can_2014}
Dürr, D., S.~Goldstein, T.~Norsen, W.~Struyve, and N.~Zanghì. 2014.
\newblock Can {Bohmian} mechanics be made relativistic?
\newblock {\em Proceedings of the Royal Society A: Mathematical, Physical and Engineering Sciences\/}~{\em 470\/}(2162): 20130699 .

\bibitem[\protect\citeauthoryear{Dürr, Goldstein, and Zanghi}{Dürr et~al.}{1992}]{durr_quantum_1992}
Dürr, D., S.~Goldstein, and N.~Zanghi. 1992.
\newblock Quantum equilibrium and the origin of absolute uncertainty.
\newblock {\em Journal of Statistical Physics\/}~{\em 67\/}(5): 843--907 .

\bibitem[\protect\citeauthoryear{Eagle}{Eagle}{2010}]{eagle_philosophy_2010}
Eagle, A. 2010.
\newblock {\em Philosophy of probability: {Contemporary} readings}.
\newblock Routlegde.

\bibitem[\protect\citeauthoryear{Everett}{Everett}{1957}]{everett_relative_1957}
Everett, H. 1957.
\newblock "relative state" formulation of {Quantum} mechanics.
\newblock {\em Rev. Mod. Phys.\/}~{\em 29\/}(3): 454--462 .

\bibitem[\protect\citeauthoryear{Everett}{Everett}{1973}]{everett_iii_theory_1973}
Everett, H. 1973.
\newblock The theory of the universal wave function, In {\em The many-worlds interpretation of quantum mechanics},  eds. DeWitt, B.S. and N.~Graham,  3--140. Princeton: Princeton University Press.

\bibitem[\protect\citeauthoryear{Feynman}{Feynman}{1951}]{feynman_concept_1951}
Feynman, R.P. 1951.
\newblock The concept of probability in quantum mechanics.
\newblock In J.~Neyman (Ed.), {\em Proceedings of the Second Berkeley Symposium on Mathematical Statistics and Probability}, Berkeley, pp.\  533--542. University of California Press.

\bibitem[\protect\citeauthoryear{Gell-Mann and Hartle}{Gell-Mann and Hartle}{1993}]{gell_classical_1993}
Gell-Mann, M. and J.B. Hartle. 1993.
\newblock Classical equations for quantum systems.
\newblock {\em Physical Review D\/}~{\em 47\/}(8): 3345 .

\bibitem[\protect\citeauthoryear{Gell-Mann and Hartle}{Gell-Mann and Hartle}{1994}]{gell1994time}
Gell-Mann, M. and J.B. Hartle. 1994.
\newblock Time symmetry and asymmetry in quantum mechanics and quantum cosmology.
\newblock {\em Physical origins of time asymmetry\/}~1: 311--345 .

\bibitem[\protect\citeauthoryear{Gell-Mann and Hartle}{Gell-Mann and Hartle}{1996}]{gell_quantum_1996}
Gell-Mann, M. and J.B. Hartle 1996.
\newblock Quantum mechanics in the light of quantum cosmology.
\newblock In {\em Foundations of Quantum Mechanics in the Light of New Technology: Selected Papers from the Proceedings of the First through Fourth International Symposia on Foundations of Quantum Mechanics}, pp.\  347--369. World Scientific.

\bibitem[\protect\citeauthoryear{Ghirardi, Rimini, and Weber}{Ghirardi et~al.}{1986}]{ghirardi_unified_1986}
Ghirardi, G.C., A.~Rimini, and T.~Weber. 1986.
\newblock Unified dynamics for microscopic and macroscopic systems.
\newblock {\em Phys. Rev. D\/}~{\em 34\/}(2): 470--491 .

\bibitem[\protect\citeauthoryear{Giere}{Giere}{1973}]{giere_objective_1973}
Giere, R.N. 1973.
\newblock Objective single-case probabilities and the foundations of statistics, {\em Studies in {Logic} and the {Foundations} of {Mathematics}}, Volume~74,  467--483. Elsevier.

\bibitem[\protect\citeauthoryear{Giovannetti, Lloyd, and Maccone}{Giovannetti et~al.}{2015}]{giovannetti_quantum_2015}
Giovannetti, V., S.~Lloyd, and L.~Maccone. 2015, August.
\newblock Quantum time.
\newblock {\em Phys. Rev. D\/}~{\em 92\/}(4): 045033.
\newblock \doi{10.1103/PhysRevD.92.045033} .

\bibitem[\protect\citeauthoryear{Giovannetti, Lloyd, and Maccone}{Giovannetti et~al.}{2023}]{giovannetti2023geometric}
Giovannetti, V., S.~Lloyd, and L.~Maccone. 2023.
\newblock Geometric event-based quantum mechanics.
\newblock {\em New Journal of Physics\/}~{\em 25\/}(2): 023027 .

\bibitem[\protect\citeauthoryear{Groisman, Hallakoun, and Vaidman}{Groisman et~al.}{2013}]{groisman_measure_2013}
Groisman, B., N.~Hallakoun, and L.~Vaidman. 2013.
\newblock The measure of existence of a quantum world and the {Sleeping} {Beauty} {Problem}.
\newblock {\em Analysis\/}~{\em 73\/}(4): 695--706 .

\bibitem[\protect\citeauthoryear{Hacking}{Hacking}{1971}]{hacking_equipossibility_1971}
Hacking, I. 1971.
\newblock Equipossibility theories of probability.
\newblock {\em The British Journal for the Philosophy of Science\/}~{\em 22\/}(4): 339--355 .

\bibitem[\protect\citeauthoryear{Hacking}{Hacking}{2006}]{hacking_emergence_2006}
Hacking, I. 2006.
\newblock {\em The emergence of probability: a philosophical study of early ideas about probability, induction and statistical inference\/} (2nd ed ed.).
\newblock Cambridge ; New York: Cambridge University Press.
\newblock OCLC: ocm68965747.

\bibitem[\protect\citeauthoryear{Harrigan and Spekkens}{Harrigan and Spekkens}{2010}]{harrigan_einstein_2010}
Harrigan, N. and R.W. Spekkens. 2010.
\newblock Einstein, incompleteness, and the epistemic view of quantum states.
\newblock {\em Foundations of Physics\/}~{\em 40\/}(2): 125--157 .

\bibitem[\protect\citeauthoryear{Hartle}{Hartle}{2021}]{hartle2021we}
Hartle, J. 2021.
\newblock What do we learn by deriving born's rule.

\bibitem[\protect\citeauthoryear{Hedden}{Hedden}{2015}]{hedden_time-slice_2015}
Hedden, B. 2015.
\newblock Time-slice rationality.
\newblock {\em Mind\/}~{\em 124\/}(494): 449--491 .

\bibitem[\protect\citeauthoryear{H{\"o}hn, Smith, and Lock}{H{\"o}hn et~al.}{2021}]{hohn2021trinity}
H{\"o}hn, P.A., A.R. Smith, and M.P. Lock. 2021.
\newblock Trinity of relational quantum dynamics.
\newblock {\em Physical Review D\/}~{\em 104\/}(6): 066001 .

\bibitem[\protect\citeauthoryear{Horwitz, Arshansky, and Elitzur}{Horwitz et~al.}{1988}]{horwitz_two_1988}
Horwitz, L., R.~Arshansky, and A.~Elitzur. 1988.
\newblock On the two aspects of time: {The} distinction and its implications.
\newblock {\em Foundations of Physics\/}~{\em 18\/}(12): 1159--1193 .

\bibitem[\protect\citeauthoryear{Huber}{Huber}{2023}]{huber2023defending}
Huber, M. 2023.
\newblock Defending many worlds via case discrimination: An attempt to showcase the conceptual incoherence of anti-realist interpretations and relational quantum mechanics.
\newblock {\em Quantum Reports\/}~{\em 5\/}(2): 345--369 .

\bibitem[\protect\citeauthoryear{Isham and Linden}{Isham and Linden}{1995}]{isham1995continuous}
Isham, C. and N.~Linden. 1995.
\newblock Continuous histories and the history group in generalized quantum theory.
\newblock {\em Journal of Mathematical Physics\/}~{\em 36\/}(10): 5392--5408 .

\bibitem[\protect\citeauthoryear{Ismael and Schaffer}{Ismael and Schaffer}{2020}]{ismael_quantum_2020}
Ismael, J. and J.~Schaffer. 2020.
\newblock Quantum holism: nonseparability as common ground.
\newblock {\em Synthese\/}~{\em 197\/}(10): 4131--4160 .

\bibitem[\protect\citeauthoryear{Ismael}{Ismael}{2009}]{ismael2009probability}
Ismael, J.T. 2009.
\newblock Probability in deterministic physics.
\newblock {\em The Journal of Philosophy\/}~{\em 106\/}(2): 89--108 .

\bibitem[\protect\citeauthoryear{Jaynes}{Jaynes}{1957}]{jaynes_information_1957}
Jaynes, E.T. 1957.
\newblock Information theory and statistical mechanics.
\newblock {\em Physical review\/}~{\em 106\/}(4): 620 .

\bibitem[\protect\citeauthoryear{Jaynes}{Jaynes}{1973}]{jaynes_well-posed_1973}
Jaynes, E.T. 1973.
\newblock The well-posed problem.
\newblock {\em Foundations of Physics\/}~{\em 3\/}(4): 477--492 .

\bibitem[\protect\citeauthoryear{Keldysh}{Keldysh}{1964}]{keldysh_diagram_1964}
Keldysh, L.V. 1964.
\newblock Diagram technique for nonequilibrium processes.
\newblock {\em Zh. Eksp. Teor. Fiz.\/}~47: 1515--1527 .

\bibitem[\protect\citeauthoryear{Kent}{Kent}{1990}]{kent_against_1990}
Kent, A. 1990.
\newblock Against many-worlds interpretations.
\newblock {\em International Journal of Modern Physics A\/}~{\em 5\/}(09): 1745--1762 .

\bibitem[\protect\citeauthoryear{Langen, Erne, Geiger, Rauer, Schweigler, Kuhnert, Rohringer, Mazets, Gasenzer, and Schmiedmayer}{Langen et~al.}{2015}]{langen2015experimental}
Langen, T., S.~Erne, R.~Geiger, B.~Rauer, T.~Schweigler, M.~Kuhnert, W.~Rohringer, I.E. Mazets, T.~Gasenzer, and J.~Schmiedmayer. 2015.
\newblock Experimental observation of a generalized gibbs ensemble.
\newblock {\em Science\/}~{\em 348\/}(6231): 207--211 .

\bibitem[\protect\citeauthoryear{Leifer}{Leifer}{2014}]{leifer_is_2014}
Leifer, M.S. 2014.
\newblock Is the quantum state real? an extended review of $\psi$-ontology theorems.

\bibitem[\protect\citeauthoryear{Lewis}{Lewis}{1976}]{lewis1976survival}
Lewis, D. 1976.
\newblock Survival and identity, In {\em The identities of persons},  ed. Rorty, A.O.,  17--40. California: University of California Press.

\bibitem[\protect\citeauthoryear{Lewis}{Lewis}{1980}]{lewis_subjectivists_1980}
Lewis, D. 1980.
\newblock A subjectivist’s guide to objective chance, {\em Ifs},  267--297. Springer.

\bibitem[\protect\citeauthoryear{Lewis}{Lewis}{2001}]{lewis_sleeping_2001}
Lewis, D. 2001.
\newblock Sleeping beauty: reply to {Elga}.
\newblock {\em Analysis\/}~{\em 61\/}(3): 171--176 .

\bibitem[\protect\citeauthoryear{Lewis}{Lewis}{2007}]{lewis2007uncertainty}
Lewis, P.J. 2007.
\newblock Uncertainty and probability for branching selves.
\newblock {\em Studies In History and Philosophy of Science Part B: Studies In History and Philosophy of Modern Physics\/}~{\em 38\/}(1): 1--14 .

\bibitem[\protect\citeauthoryear{Lewis}{Lewis}{2009}]{lewis2009probability}
Lewis, P.J. 2009.
\newblock Probability, self-location, and quantum branching.
\newblock {\em Philosophy of Science\/}~{\em 76\/}(5): 1009--1019 .

\bibitem[\protect\citeauthoryear{Maccone}{Maccone}{2019}]{maccone_fundamental_2019}
Maccone, L. 2019.
\newblock A fundamental problem in quantizing general relativity.
\newblock {\em Foundations of Physics\/}~{\em 49\/}(12): 1394--1403 .

\bibitem[\protect\citeauthoryear{Maccone and Sacha}{Maccone and Sacha}{2020}]{maccone_quantum_2020}
Maccone, L. and K.~Sacha. 2020.
\newblock Quantum measurements of time.
\newblock {\em Physical Review Letters\/}~{\em 124\/}(11): 110402 .

\bibitem[\protect\citeauthoryear{Maghsoudi and Taheri~Khorramabadi}{Maghsoudi and Taheri~Khorramabadi}{2024}]{maghsoudi2024evolutionary}
Maghsoudi, M.E. and S.A. Taheri~Khorramabadi. 2024.
\newblock The evolutionary versus the all-at-once picture of spacetime.
\newblock {\em Foundations of Physics\/}~{\em 54\/}(5): 64 .

\bibitem[\protect\citeauthoryear{Mandolesi}{Mandolesi}{2019}]{mandolesi2019analysis}
Mandolesi, A.L. 2019.
\newblock Analysis of wallace’s proof of the born rule in everettian quantum mechanics ii: concepts and axioms.
\newblock {\em Foundations of Physics\/}~{\em 49\/}(1): 24--52 .

\bibitem[\protect\citeauthoryear{Mandolesi}{Mandolesi}{2020}]{mandolesi_quantum_2020}
Mandolesi, A.L. 2020.
\newblock Quantum fractionalism: {The} {Born} rule as a consequence of the complex {Pythagorean} theorem.
\newblock {\em Physics Letters A\/}~{\em 384\/}(28): 126725 .

\bibitem[\protect\citeauthoryear{Marchildon}{Marchildon}{2015}]{marchildon_multiplicity_2015}
Marchildon, L. 2015.
\newblock Multiplicity in everett's interpretation of quantum mechanics.
\newblock {\em Studies in History and Philosophy of Science Part B: Studies in History and Philosophy of Modern Physics\/}~52: 274--284 .

\bibitem[\protect\citeauthoryear{Maudlin}{Maudlin}{1995}]{maudlin_three_1995}
Maudlin, T. 1995.
\newblock Three measurement problems.
\newblock {\em Topoi\/}~{\em 14\/}(1): 7--15 .

\bibitem[\protect\citeauthoryear{Maudlin}{Maudlin}{2007}]{maudlin_what_2007}
Maudlin, T. 2007.
\newblock What could be objective about probabilities?
\newblock {\em Studies in History and Philosophy of Science Part B: Studies in History and Philosophy of Modern Physics\/}~{\em 38\/}(2): 275--291 .

\bibitem[\protect\citeauthoryear{McQueen and Vaidman}{McQueen and Vaidman}{2019}]{mcqueen_defence_2019}
McQueen, K.J. and L.~Vaidman. 2019, May.
\newblock In defence of the self-location uncertainty account of probability in the many-worlds interpretation.
\newblock {\em Studies in History and Philosophy of Science Part B: Studies in History and Philosophy of Modern Physics\/}~66: 14--23 .

\bibitem[\protect\citeauthoryear{Megidish, Halevy, Shacham, Dvir, Dovrat, and Eisenberg}{Megidish et~al.}{2013}]{megidish2013entanglement}
Megidish, E., A.~Halevy, T.~Shacham, T.~Dvir, L.~Dovrat, and H.~Eisenberg. 2013.
\newblock Entanglement swapping between photons that have never coexisted.
\newblock {\em Physical review letters\/}~{\em 110\/}(21): 210403 .

\bibitem[\protect\citeauthoryear{Miller}{Miller}{2016}]{miller_quantum_2016}
Miller, E. 2016.
\newblock Quantum holism.
\newblock {\em Philosophy compass\/}~{\em 11\/}(9): 507--514 .

\bibitem[\protect\citeauthoryear{Nowakowski, Cohen, and Horodecki}{Nowakowski et~al.}{2018}]{nowakowski2018entangled}
Nowakowski, M., E.~Cohen, and P.~Horodecki. 2018.
\newblock Entangled histories versus the two-state-vector formalism: Towards a better understanding of quantum temporal correlations.
\newblock {\em Physical Review A\/}~{\em 98\/}(3): 032312 .

\bibitem[\protect\citeauthoryear{Page and Wootters}{Page and Wootters}{1983}]{page_evolution_1983}
Page, D.N. and W.K. Wootters. 1983.
\newblock Evolution without evolution: {Dynamics} described by stationary observables.
\newblock {\em Physical Review D\/}~{\em 27\/}(12): 2885 .

\bibitem[\protect\citeauthoryear{Papineau}{Papineau}{2010}]{papineau_fair_2010}
Papineau, D. 2010.
\newblock A fair deal for everettians, {\em Many worlds?: Everett, quantum theory, \& reality},  206--226. Oxford University Press.

\bibitem[\protect\citeauthoryear{Parker and Jeynes}{Parker and Jeynes}{2023}]{parker2023maximum}
Parker, M.C. and C.~Jeynes. 2023.
\newblock A maximum entropy resolution to the wine/water paradox.
\newblock {\em Entropy\/}~{\em 25\/}(8): 1242 .

\bibitem[\protect\citeauthoryear{Pearle}{Pearle}{1989}]{pearle_combining_1989}
Pearle, P. 1989.
\newblock Combining stochastic dynamical state-vector reduction with spontaneous localization.
\newblock {\em Phys. Rev. A\/}~{\em 39\/}(5): 2277--2289 .

\bibitem[\protect\citeauthoryear{Penrose}{Penrose}{1998}]{penrose1998quantum}
Penrose, R. 1998.
\newblock Quantum computation, entanglement and state reduction.
\newblock {\em Philosophical Transactions of the Royal Society of London. Series A: Mathematical, Physical and Engineering Sciences\/}~{\em 356\/}(1743): 1927--1939 .

\bibitem[\protect\citeauthoryear{Pitowsky}{Pitowsky}{2006}]{pitowsky_quantum_2006}
Pitowsky, I. 2006.
\newblock Quantum mechanics as a theory of probability, {\em Physical theory and its interpretation},  213--240. Springer.

\bibitem[\protect\citeauthoryear{Popper}{Popper}{1959}]{popper_propensity_1959}
Popper, K.R. 1959.
\newblock The propensity interpretation of probability.
\newblock {\em The British journal for the philosophy of science\/}~{\em 10\/}(37): 25--42 .

\bibitem[\protect\citeauthoryear{Pusey, Barrett, and Rudolph}{Pusey et~al.}{2012}]{pusey_reality_2012}
Pusey, M.F., J.~Barrett, and T.~Rudolph. 2012.
\newblock On the reality of the quantum state.
\newblock {\em Nature Phys\/}~{\em 8\/}(6): 475--478 .

\bibitem[\protect\citeauthoryear{Ramsey}{Ramsey}{1926}]{ramsey1926truth}
Ramsey, F.P. 1926.
\newblock Truth and probability, {\em Readings in Formal Epistemology: Sourcebook},  21--45. Springer.

\bibitem[\protect\citeauthoryear{Ridley}{Ridley}{2023}]{ridley2023quantum}
Ridley, M. 2023.
\newblock Quantum probability from temporal structure.
\newblock {\em Quantum Reports\/}~{\em 5\/}(2): 496--509 .

\bibitem[\protect\citeauthoryear{Ridley and Adlam}{Ridley and Adlam}{2025}]{ridley2025time}
Ridley, M. and E.~Adlam. 2025.
\newblock Time and event symmetry in quantum mechanics.
\newblock {\em Quantum Studies: Mathematics and Foundations\/}~{\em 12\/}(1): 9 .

\bibitem[\protect\citeauthoryear{Riedel, Zurek, and Zwolak}{Riedel et~al.}{2016}]{riedel_objective_2016}
Riedel, C.J., W.H. Zurek, and M.~Zwolak. 2016.
\newblock Objective past of a quantum universe: {Redundant} records of consistent histories.
\newblock {\em Physical Review A\/}~{\em 93\/}(3): 032126 .

\bibitem[\protect\citeauthoryear{Rovelli}{Rovelli}{1998}]{rovelli_incerto_1998}
Rovelli, C. 1998.
\newblock “{Incerto} tempore, incertisque loci”: {Can} we compute the exact time at which a quantum measurement happens?
\newblock {\em Foundations of Physics\/}~{\em 28\/}(7): 1031--1043 .

\bibitem[\protect\citeauthoryear{Saunders}{Saunders}{1993}]{saunders_decoherence_1993}
Saunders, S. 1993.
\newblock Decoherence, relative states, and evolutionary adaptation.
\newblock {\em Foundations of physics\/}~{\em 23\/}(12): 1553--1585 .

\bibitem[\protect\citeauthoryear{Saunders}{Saunders}{1995}]{saunders1995time}
Saunders, S. 1995.
\newblock Time, quantum mechanics, and decoherence.
\newblock {\em Synthese\/}~{\em 102\/}(2): 235--266 .

\bibitem[\protect\citeauthoryear{Saunders}{Saunders}{2021a}]{saunders_branch-counting_2021}
Saunders, S. 2021a.
\newblock Branch-counting in the {Everett} interpretation of quantum mechanics.
\newblock {\em Proceedings of the Royal Society A\/}~{\em 477\/}(2255): 20210600 .

\bibitem[\protect\citeauthoryear{Saunders}{Saunders}{2021b}]{saunders2021everett}
Saunders, S. 2021b.
\newblock The everett interpretation: Probability 1, {\em The Routledge Companion to Philosophy of Physics},  230--246. Routledge.

\bibitem[\protect\citeauthoryear{Saunders, Barrett, Kent, and Wallace}{Saunders et~al.}{2010}]{saunders_many_2010}
Saunders, S., J.~Barrett, A.~Kent, and D.~Wallace. 2010.
\newblock {\em Many worlds?: {Everett}, quantum theory, \& reality}.
\newblock Oxford University Press.

\bibitem[\protect\citeauthoryear{Saunders and Wallace}{Saunders and Wallace}{2008}]{saunders_branching_2008}
Saunders, S. and D.~Wallace. 2008.
\newblock Branching and uncertainty.
\newblock {\em The British Journal for the Philosophy of Science\/}~{\em 59\/}(3): 293--305 .

\bibitem[\protect\citeauthoryear{Schleder, Padilha, Acosta, Costa, and Fazzio}{Schleder et~al.}{2019}]{schleder2019dft}
Schleder, G.R., A.C. Padilha, C.M. Acosta, M.~Costa, and A.~Fazzio. 2019.
\newblock From dft to machine learning: recent approaches to materials science--a review.
\newblock {\em Journal of Physics: Materials\/}~{\em 2\/}(3): 032001 .

\bibitem[\protect\citeauthoryear{Schr{\"o}dinger}{Schr{\"o}dinger}{1935}]{schrodinger1935discussion}
Schr{\"o}dinger, E. 1935.
\newblock Discussion of probability relations between separated systems.
\newblock {\em Mathematical Proceedings of the Cambridge Philosophical Society\/}~{\em 31\/}(4): 555--563 .

\bibitem[\protect\citeauthoryear{Schwinger}{Schwinger}{1961}]{schwinger1961brownian}
Schwinger, J. 1961.
\newblock Brownian motion of a quantum oscillator.
\newblock {\em Journal of Mathematical Physics\/}~{\em 2\/}(3): 407--432 .

\bibitem[\protect\citeauthoryear{Sebens and Carroll}{Sebens and Carroll}{2018}]{sebens_self-locating_2018}
Sebens, C.T. and S.M. Carroll. 2018.
\newblock Self-locating uncertainty and the origin of probability in {Everettian} quantum mechanics.
\newblock {\em The British Journal for the Philosophy of Science\/}~{\em 69\/}(1): 25--74 .

\bibitem[\protect\citeauthoryear{Shankar}{Shankar}{2012}]{shankar2012principles}
Shankar, R. 2012.
\newblock {\em Principles of quantum mechanics}.
\newblock Springer Science \& Business Media.

\bibitem[\protect\citeauthoryear{Short}{Short}{2023}]{short2023probability}
Short, A.J. 2023.
\newblock Probability in many-worlds theories.
\newblock {\em Quantum\/}~7: 971 .

\bibitem[\protect\citeauthoryear{Silva, Guryanova, Short, Skrzypczyk, Brunner, and Popescu}{Silva et~al.}{2017}]{silva_connecting_2017}
Silva, R., Y.~Guryanova, A.J. Short, P.~Skrzypczyk, N.~Brunner, and S.~Popescu. 2017.
\newblock Connecting processes with indefinite causal order and multi-time quantum states.
\newblock {\em New Journal of Physics\/}~{\em 19\/}(10): 103022 .

\bibitem[\protect\citeauthoryear{Spekkens}{Spekkens}{2007}]{spekkens_evidence_2007}
Spekkens, R.W. 2007.
\newblock Evidence for the epistemic view of quantum states: A toy theory.
\newblock {\em Phys. Rev. A\/}~{\em 75\/}(3): 032110 .

\bibitem[\protect\citeauthoryear{Stefanucci and Van~Leeuwen}{Stefanucci and Van~Leeuwen}{2025}]{stefanucci2025nonequilibrium}
Stefanucci, G. and R.~Van~Leeuwen. 2025.
\newblock {\em Nonequilibrium many-body theory of quantum systems: a modern introduction, Second Edition}.
\newblock Cambridge University Press.

\bibitem[\protect\citeauthoryear{Stoica}{Stoica}{2021}]{stoica2021post}
Stoica, O.C. 2021.
\newblock The post-determined block universe.
\newblock {\em Quantum Studies: Mathematics and Foundations\/}~{\em 8\/}(1): 69--101 .

\bibitem[\protect\citeauthoryear{Strasberg, Reinhard, and Schindler}{Strasberg et~al.}{2024}]{strasberg2024}
Strasberg, P., T.E. Reinhard, and J.~Schindler. 2024.
\newblock First principles numerical demonstration of emergent decoherent histories.
\newblock {\em Phys. Rev. X\/}~14: 041027 .

\bibitem[\protect\citeauthoryear{Strasberg and Schindler}{Strasberg and Schindler}{2023}]{strasberg2023shearing}
Strasberg, P. and J.~Schindler. 2023.
\newblock Shearing off the tree: Emerging branch structure and born's rule in the multiverse.

\bibitem[\protect\citeauthoryear{Strevens}{Strevens}{1999}]{strevens_objective_1999}
Strevens, M. 1999.
\newblock Objective probability as a guide to the world.
\newblock {\em Philosophical Studies\/}~{\em 95\/}(3): 243--275 .

\bibitem[\protect\citeauthoryear{Susskind}{Susskind}{2016}]{susskind_copenhagen_2016}
Susskind, L. 2016.
\newblock Copenhagen vs {Everett}, teleportation, and {ER}= {EPR}.
\newblock {\em Fortschritte der Physik\/}~{\em 64\/}(6-7): 551--564 .

\bibitem[\protect\citeauthoryear{Tappenden}{Tappenden}{2011}]{tappenden_evidence_2011}
Tappenden, P. 2011, March.
\newblock Evidence and {Uncertainty} in {Everett}'s {Multiverse}.
\newblock {\em The British Journal for the Philosophy of Science\/}~{\em 62\/}(1): 99--123 .

\bibitem[\protect\citeauthoryear{Taylor, Cheung, Brukner, and Vedral}{Taylor et~al.}{2004}]{taylor2004entanglement}
Taylor, S., S.~Cheung, {\v{C}}.~Brukner, and V.~Vedral. 2004.
\newblock Entanglement in time and temporal communication complexity.
\newblock {\em AIP Conference Proceedings\/}~{\em 734\/}(1): 281--284 .

\bibitem[\protect\citeauthoryear{Tumulka}{Tumulka}{2006}]{tumulka_relativistic_2006}
Tumulka, R. 2006.
\newblock A relativistic version of the {Ghirardi}–{Rimini}–{Weber} model.
\newblock {\em Journal of Statistical Physics\/}~{\em 125\/}(4): 821--840 .

\bibitem[\protect\citeauthoryear{Vaidman}{Vaidman}{1987}]{vaidman_problem_1987}
Vaidman, L. 1987.
\newblock {\em The problem of the interpretation of relativistic quantum theories}.
\newblock Ph.\ D. thesis, Tel-Aviv University.

\bibitem[\protect\citeauthoryear{Vaidman}{Vaidman}{1998}]{vaidman_schizophrenic_1998}
Vaidman, L. 1998, October.
\newblock On schizophrenic experiences of the neutron or why we should believe in the many‐worlds interpretation of quantum theory.
\newblock {\em International Studies in the Philosophy of Science\/}~{\em 12\/}(3): 245--261 .

\bibitem[\protect\citeauthoryear{Vaidman}{Vaidman}{2012}]{vaidman_probability_2012}
Vaidman, L. 2012.
\newblock Probability in the many-worlds interpretation of quantum mechanics, {\em Probability in physics},  299--311. Springer.

\bibitem[\protect\citeauthoryear{Vaidman}{Vaidman}{2014}]{vaidman_quantum_2014}
Vaidman, L. 2014.
\newblock Quantum theory and determinism.
\newblock {\em Quantum Studies: Mathematics and Foundations\/}~{\em 1\/}(1-2): 5--38 .

\bibitem[\protect\citeauthoryear{Vaidman}{Vaidman}{2016}]{vaidman2016all}
Vaidman, L. 2016.
\newblock All is $\psi$.
\newblock {\em Journal of Physics: Conference Series\/}~{\em 701\/}(1): 012020 .

\bibitem[\protect\citeauthoryear{Vaidman}{Vaidman}{2020}]{vaidman_derivations_2020}
Vaidman, L. 2020.
\newblock Derivations of the {Born} rule, {\em Quantum, {Probability}, {Logic}},  567--584. Springer.

\bibitem[\protect\citeauthoryear{Vaidman}{Vaidman}{2021}]{vaidman_many-worlds_2021}
Vaidman, L. 2021.
\newblock Many-worlds interpretation of quantum mechanics.

\bibitem[\protect\citeauthoryear{Vaidman}{Vaidman}{2022}]{vaidman2022many}
Vaidman, L. 2022.
\newblock Why the many-worlds interpretation?
\newblock {\em Quantum Reports\/}~{\em 4\/}(3): 264--271 .

\bibitem[\protect\citeauthoryear{Van~Fraassen}{Van~Fraassen}{1989}]{van_fraassen_laws_1989}
Van~Fraassen, B.C. 1989.
\newblock {\em Laws and symmetry}.
\newblock Clarendon Press.

\bibitem[\protect\citeauthoryear{von Mises}{von Mises}{1939}]{mises_probability_1939}
von Mises, R. 1939.
\newblock {\em Probability, statistics and truth.}
\newblock Macmillan.

\bibitem[\protect\citeauthoryear{Wallace}{Wallace}{2002}]{wallace_worlds_2002}
Wallace, D. 2002.
\newblock Worlds in the {Everett} interpretation.
\newblock {\em Studies in History and Philosophy of Science Part B: Studies in History and Philosophy of Modern Physics\/}~{\em 33\/}(4): 637--661 .

\bibitem[\protect\citeauthoryear{Wallace}{Wallace}{2010a}]{wallace_decoherence_2010}
Wallace, D. 2010a.
\newblock Decoherence and ontology: Or: How i learned to stop worrying and love fapp, {\em Many worlds?: Everett, quantum theory, \& reality},  53--72. Oxford University Press.

\bibitem[\protect\citeauthoryear{Wallace}{Wallace}{2010b}]{wallace_how_2010}
Wallace, D. 2010b.
\newblock How to prove the {Born} rule, {\em Many worlds?: Everett, quantum theory, \& reality},  227--263. Oxford University Press.

\bibitem[\protect\citeauthoryear{Wallace}{Wallace}{2012}]{wallace_emergent_2012}
Wallace, D. 2012.
\newblock {\em The emergent multiverse: {Quantum} theory according to the {Everett} interpretation}.
\newblock Oxford University Press.

\bibitem[\protect\citeauthoryear{Wallace}{Wallace}{2020}]{wallace_plurality_2020}
Wallace, D. 2020.
\newblock On the plurality of quantum theories: Quantum theory as a framework, and its implications for the quantum measurement problem, {\em Scientific Realism and the Quantum}, ~78. Oxford University Press.

\bibitem[\protect\citeauthoryear{Wallace and Timpson}{Wallace and Timpson}{2010}]{wallace_quantum_2010}
Wallace, D. and C.G. Timpson. 2010.
\newblock Quantum mechanics on spacetime {I}: {Spacetime} state realism.
\newblock {\em The British journal for the philosophy of science\/}~{\em 61\/}(4): 697--727 .

\bibitem[\protect\citeauthoryear{Wharton}{Wharton}{2015}]{wharton2015universe}
Wharton, K. 2015.
\newblock The universe is not a computer, {\em Questioning the Foundations of Physics: Which of Our Fundamental Assumptions Are Wrong?},  177--189. Springer.

\bibitem[\protect\citeauthoryear{Wharton and Argaman}{Wharton and Argaman}{2020}]{wharton2020colloquium}
Wharton, K.B. and N.~Argaman. 2020.
\newblock Colloquium: Bell’s theorem and locally mediated reformulations of quantum mechanics.
\newblock {\em Reviews of Modern Physics\/}~{\em 92\/}(2): 021002 .

\bibitem[\protect\citeauthoryear{Wilhelm}{Wilhelm}{2021}]{wilhelm2021centering}
Wilhelm, I. 2021.
\newblock Centering the principal principle.
\newblock {\em Philosophical Studies\/}~{\em 178\/}(6): 1897--1915 .

\bibitem[\protect\citeauthoryear{Wilhelm}{Wilhelm}{2022}]{wilhelm2022centering}
Wilhelm, I. 2022.
\newblock Centering the everett interpretation.
\newblock {\em The Philosophical Quarterly\/}~{\em 72\/}(4): 1019--1039 .

\bibitem[\protect\citeauthoryear{Wilhelm}{Wilhelm}{2023}]{wilhelm2023centering}
Wilhelm, I. 2023.
\newblock Centering the born rule.
\newblock {\em Quantum Reports\/}~{\em 5\/}(1): 311--324 .

\bibitem[\protect\citeauthoryear{Williamson}{Williamson}{2010}]{williamson_defence_2010}
Williamson, J. 2010.
\newblock {\em In defence of objective {Bayesianism}}, Volume~24.
\newblock Oxford University Press.

\bibitem[\protect\citeauthoryear{Wils}{Wils}{1970}]{wils1970direct}
Wils, W. 1970.
\newblock Direct integrals of hilbert spaces i.
\newblock {\em Mathematica Scandinavica\/}~{\em 26\/}(1): 73--88 .

\bibitem[\protect\citeauthoryear{Wilson}{Wilson}{2012}]{wilson_everettian_2012}
Wilson, A. 2012.
\newblock Everettian quantum mechanics without branching time.
\newblock {\em Synthese\/}~{\em 188\/}(1): 67--84 .

\bibitem[\protect\citeauthoryear{Wilson}{Wilson}{2013}]{wilson_objective_2013}
Wilson, A. 2013.
\newblock Objective probability in {Everettian} quantum mechanics.
\newblock {\em The British Journal for the Philosophy of Science\/}~{\em 64\/}(4): 709--737 .

\bibitem[\protect\citeauthoryear{Zurek}{Zurek}{1993}]{zurek1993preferred}
Zurek, W.H. 1993.
\newblock Preferred states, predictability, classicality and the environment-induced decoherence.
\newblock {\em Progress of Theoretical Physics\/}~{\em 89\/}(2): 281--312 .

\bibitem[\protect\citeauthoryear{Zurek}{Zurek}{2003a}]{zurek2003decoherence}
Zurek, W.H. 2003a.
\newblock Decoherence, einselection, and the quantum origins of the classical.
\newblock {\em Reviews of modern physics\/}~{\em 75\/}(3): 715 .

\bibitem[\protect\citeauthoryear{Zurek}{Zurek}{2003b}]{zurek_environment-assisted_2003}
Zurek, W.H. 2003b.
\newblock Environment-assisted invariance, entanglement, and probabilities in quantum physics.
\newblock {\em Physical review letters\/}~{\em 90\/}(12): 120404 .

\bibitem[\protect\citeauthoryear{Zurek}{Zurek}{2005}]{zurek_probabilities_2005}
Zurek, W.H. 2005.
\newblock Probabilities from entanglement, {Born}’s rule $p_{k} = \left| \psi_k \right|^2$ from envariance.
\newblock {\em Physical Review A\/}~{\em 71\/}(5): 052105 .

\bibitem[\protect\citeauthoryear{Zurek}{Zurek}{2007}]{zurek_relative_2007}
Zurek, W.H. 2007.
\newblock Relative states and the environment: einselection, envariance, quantum darwinism, and the existential interpretation.

\bibitem[\protect\citeauthoryear{Zurek}{Zurek}{2018}]{zurek_quantum_2018}
Zurek, W.H. 2018.
\newblock Quantum theory of the classical: quantum jumps, {Born}’s {Rule} and objective classical reality via quantum {Darwinism}.
\newblock {\em Philosophical Transactions of the Royal Society A: Mathematical, Physical and Engineering Sciences\/}~{\em 376\/}(2123): 20180107 .

\bibitem[\protect\citeauthoryear{Zurek}{Zurek}{2022}]{zurek2022quantum}
Zurek, W.H. 2022.
\newblock Quantum theory of the classical: Einselection, envariance, quantum darwinism and extantons.
\newblock {\em Entropy\/}~{\em 24\/}(11): 1520 .

\end{thebibliography}
\end{document}